\documentclass[article,superscriptaddress,showpacs,floatfix,nofootinbib,twocolumn]{revtex4-1}
\usepackage{amsmath,graphicx,float,upgreek,mciteplus,titlesec,appendix}
\renewcommand{\thesection}{\Roman{section}} 
\usepackage{mathtools}
\usepackage{scalerel} 

\titleformat{\section}
[hang]
{\normalsize\bfseries}
{\thesection.}
{1em}
{\MakeUppercase}
\titlespacing{\section}
{10pt}
{15pt}
{8pt}

\titleformat{\subsection}
[hang]
{\bfseries\normalsize\centering}
{\thesubsection.}
{1em}
{}
\titlespacing{\subsection}
{0pt}
{10pt}
{5pt}

\def\eqref#1{{Eq.\!~(\ref{#1})}}
\def\figref#1{{Fig.\!~\ref{#1}}}
\def\secref#1{{Sec.\!~\ref{#1}}}

\newcommand{\nr}[1]{(\ref{#1})}

\usepackage{subfigure}

\usepackage{xcolor}
\definecolor{lcolor}{rgb}{0.5,0,0}
\definecolor{citcolor}{rgb}{0,0.3,0.0}
\usepackage[breaklinks,colorlinks,urlcolor=blue,citecolor=citcolor,linkcolor=lcolor]{hyperref}

\begin{document}

\title{Evolution of initial stage fluctuations in the glasma}

\author{Pablo Guerrero-Rodr\'iguez}
\affiliation{Department  of  Physics,  University  of  Jyv\"askyl\"a  P.O.\  Box  35,  40014  University  of  Jyv\"askyl\"a,  Finland}
\affiliation{Helsinki  Institute  of  Physics,  P.O.\  Box  64,  00014  University  of  Helsinki,  Finland}
\author{Tuomas Lappi}
\affiliation{Department  of  Physics,  University  of  Jyv\"askyl\"a  P.O.\  Box  35,  40014  University  of  Jyv\"askyl\"a,  Finland}
\affiliation{Helsinki  Institute  of  Physics,  P.O.\  Box  64,  00014  University  of  Helsinki,  Finland}

\begin{abstract}
We perform a calculation of the one- and two-point correlation functions of energy density and axial charge deposited in the glasma in the initial stage of a heavy ion collision at finite proper time. We do this by describing the initial stage of heavy ion collisions in terms of freely evolving classical fields whose dynamics obey the linearized Yang-Mills equations.
Our approach allows us to systematically resum the contributions of high momentum modes that would make a power series expansion in proper time divergent.
We evaluate the field correlators in the McLerran-Venugopalan model using the glasma graph approximation, but our approach for the time dependence can be applied to a general four-point function of the initial color fields.
Our results provide analytical insight into the preequilibrium phase of heavy ion collisions without requiring a numerical solution to the Yang-Mills equations.
\end{abstract}

\maketitle

\section{Introduction}

Heavy ion collisions (HICs) open an experimental window to the most extreme regimes of quantum chromodynamics (QCD). More specifically, they give access to the quark gluon plasma (QGP), an extremely hot and dense phase of matter made up of deconfined partons. The heavy ion physics programs at the Relativistic Heavy Ion Collider and the Large Hadron Collider are devoted to making precision measurements of QGP properties, such as its collective fluidlike behavior. This feature manifests itself experimentally through nontrivial correlations in the final state of the collision. However, such correlations are reflective not only of the properties of the QGP, but also of the fluctuations of energy and momentum deposited during the earliest stages of HICs---before the QGP is formed (see e.g.\ \cite{Luzum:2013yya}).
The description of these fluctuations requires  a  degree of phenomenological modeling to incorporate contributions from different dynamical mechanisms (randomly changing nucleon positions, fluctuations of partonic degrees of freedom, color charge density fluctuations) whose relative importance is not yet very clear. A further source of uncertainty is understanding the time evolution of these fluctuations in the very early preequilibrium stage, before hydrodynamics or even kinetic theory is applicable. Addressing this time development is the primary purpose of this paper. 

Despite the inherent nonperturbative character of the dynamics in the early stages of a heavy ion collision, the color glass condensate (CGC) effective theory (see e.g.\ \cite{Gelis:2010nm,Lappi:2010ek,Albacete:2014fwa} for reviews) provides a formalism to analyze it in the weak coupling limit.  In this framework, the high density of soft (small-$x$) partons carried by the colliding nuclei is described in terms of gluon fields whose dynamics obey the classical Yang-Mills equations. This approach is valid at high energies, where the occupation numbers of small-$x$ gluons (more specifically, those with transverse momentum of the order of the saturation scale $Q_s$) are large. This justifies the use of a classical approximation, which makes it possible to both include nonlinear saturation effects, and address the time development of the system in an explicit calculation. The source of the gluon fields is the large-$x$ partons, represented as a collection of SU($N_c$) charges. The fluctuations are implemented in this context by describing the density of color charges as a stochastic quantity that varies on an event-by-event basis. In the McLerran-Venugopalan (MV) model~\cite{McLerran:1993ni,McLerran:1993ka,McLerran:1994vd}, it is assumed that these random variations obey Gaussian statistics. In this work, we will adopt the generalized Gaussian approach, where correlators of the color charges are expressed in terms of a two-point function, which needs not, however, have precisely the form of the two-point function of the original MV model. 

In the CGC picture, the multiple interactions ensuing after a HIC give rise to a coherent, highly dense substance known as glasma~\cite{Lappi:2006fp}. Within a short time, this transient state is believed to undergo an evolution process that leads the system to local thermal equilibrium. This transition is known as thermalization, and its precise theoretical characterization remains one of the most fundamental open problems of the field \cite{Dusling:2010rm,Gelis:2015gza,Berges:2020fwq,Kurkela:2016vts}.
The CGC formalism is well suited to describe the earliest phase of this evolution, during which the gluon density is large enough to warrant a classical treatment. However, as the system expands and becomes dilute, the classical field description starts to break down. A series of recent works~\cite{Kurkela:2014tea,Kurkela:2015qoa,Keegan:2016cpi,Kurkela:2018vqr} argues that an effective kinetic theory might provide the intermediate step that matches the classical description with hydrodynamics, which govern the subsequent evolution of the QGP phase. Although this discussion is of fundamental interest in the field, it is out of the scope of this paper.

Rather, in this work, we focus on the evolution of the glasma state  at even earlier times and its impact on the primordial fluctuations.
Quite generally, the magnitude of such fluctuations is encoded in the following difference of correlators:
\begin{align}
S_f(x_{\perp},y_{\perp})=\langle f(x_{\perp}) f(y_{\perp})\rangle - \langle f(x_{\perp})\rangle\langle f(y_{\perp})\rangle,\label{2pfgen}
\end{align}
where $f(x_{\perp})$ denotes the value of a property of the glasma at a point $x_{\perp}$ of the transverse plane and the notation $\langle...\rangle$ represents an average over the background fields. Such correlations have been computed analytically in previous works for both the energy density \cite{Lappi:2006hq,Lappi:2017skr,Albacete:2018bbv} and the axial charge \cite{Lappi:2006fp,Lappi:2017skr,Bhalerao:2019uzw} deposited by the nuclei right after the collision (i.e.\ for an infinitesimal positive proper time $\tau\!=\!0^+$). In the case of the energy density, this calculation has found a practical application in the construction of a model of initial conditions for HICs that was used in the description of eccentricity fluctuations in Refs.~\cite{Giacalone:2019kgg,Bhalerao:2019uzw,Gelis:2019vzt,Giacalone:2019vwh}.
Similarly, it was proposed that the two-point function of the divergence of the Chern-Simons current might be applied in the modeling of initial conditions of anomalous hydrodynamical simulations \cite{Lappi:2017skr}. These and other potential applications are founded on the assumption that, at least up until proper times $\tau\!\sim\!1/Q_s$, the thermalization process does not induce a significant modification of the fluctuations of the relevant properties. 
While this assumption is supported by studies in the case of eccentricities \cite{Schenke:2012hg}, it is uncertain whether the same can be said about their fluctuations, whose computation implies the integration of the energy density two-point function over the transverse plane. It is thus essential to have a theoretically robust notion of the evolution of this object at later times.

The evolution of the glasma is encoded in the classical Yang-Mills equations, for which no analytical solution has been found so far. A full nonperturbative solution of the time dependence, such as in the impact parameter-dependent glasma (IP-glasma) model~\cite{Schenke:2012wb,Schenke:2012hg,Mantysaari:2017cni} (see also \cite{Jia_2021} for a more recent study), relies on a numerical lattice calculation~\cite{Krasnitz:1998ns,Krasnitz:2001qu,Krasnitz:2001qu} to evolve the system from $\tau\!=\!0^+$ (where analytical solutions can be obtained and used as boundary conditions to the evolution) up to a time when the solution is matched to kinetic theory or hydrodynamics.   As will be detailed in \secref{evol}, some analytical  approaches are based on approximations such as the weak field limit, where one performs a systematic expansion in orders of the small color source densities \cite{Kovner:1995ts,Kovchegov:1997ke,Dumitru:2001ux,McLerran:2016snu,Lappi:2017skr}.
The downside of the weak field limit is that it makes the results applicable only in the dilute-dense regime. Another strategy, first proposed in \cite{Fries:2006pv} and further developed in \cite{Fujii:2008km,Chen:2015wia,Carrington:2020ssh}, is based on a systematic expansion of the gluon fields in powers of $\tau$. This expansion turns the Yang-Mills equations into an infinite system of differential equations that can be solved recursively. However, this is a series  around the  point $\tau=0$ where the solution in the MV model is not analytical~\cite{Lappi:2006hq}, and the resulting terms are found to be plagued by UV divergences. 

Here we will follow the approach also proposed in Ref.~\cite{Fujii:2008km}, where it is argued that using the Abelian (linearized) version of the Yang-Mills effectively resums the singularities, making the series finite. Remarkably, this resummation ansatz achieves the same functional $\tau$ dependence found in previous works without resorting to the weak field limit. This would in principle come at the cost of the validity range in $\tau$, which according to  Ref.~\cite{Fujii:2008km} would be reduced to very short proper times after the collision\footnote{In Ref.~\cite{Fujii:2008km}, it is also proposed that this situation can be improved by considering nonlinear corrections to the linearized Yang-Mills equations.} ($\tau\!\ll\!1/Q_s$). However, we argue that this limitation affects only the low momentum modes, which are less important for the discussion presented in this work (we focus our study on largely UV-dominated objects---the energy density and the divergence of the Chern-Simons current). Treating the dense-dense case by means of the linearized Yang-Mills equations allows us to propagate the full initial conditions of the collision (i.e.\ at $\tau\!=\!0^+$) instead of taking the first orders of an expansion of the Wilson lines in powers of color charge densities. As we will discuss (see also~\cite{Blaizot:2010kh}), this approach introduces some gauge dependence that is not there for a systematical expansion in a small source density. However, for a relatively UV-dominated quantity such as the energy density, both this gauge dependence and the screening effects caused by nonlinearities in the final state~\cite{Boguslavski:2021buh,Boguslavski:2019fsb} should be small. In combining the nonlinear initial condition with a linearized time evolution, we preserve the full gluon saturation features of the system at the initial condition. 

The paper is organized as follows. In \secref{inco}, we briefly discuss the basic elements of the glasma fields at $\tau=0^+$, as well as the specific approximations applied in this work. We also devote a part of this section to review the calculation of the one- and two-point functions of the energy density and the divergence of the Chern-Simons current of the glasma at $\tau\!=\!0^+$. In \secref{evol}, we detail our calculation of the $\tau$ evolution of these correlators. In the final section, we present our conclusions, as well as potential applications and continuations of this work. A short description of the divergence of the Chern-Simons current and its role in the search for chiral magnetic effect (CME) signals is included in Appendix \ref{CME}, and a more general result for the two-point correlators in Appendix~\ref{2pedgg}. Last, in Appendix~\ref{norc}, we briefly discuss the results obtained under the MV model with a fixed coupling constant.

\section{Initial conditions}\label{inco}

In the following section, we briefly revisit the calculation of one- and two-point correlators at $\tau\!=\!0^+$. The results described here are well established in the literature \cite{Lappi:2006hq,Lappi:2006fp,Lappi:2017skr,Albacete:2018bbv,Bhalerao:2019uzw}. However, reviewing them serves the twofold purpose of showing a fuller picture of glasma evolution and fixing notations that will be used throughout the paper.
First, we outline the basic elements of our framework.

\subsection{The MV model}\label{incoMV}
This model, presented in \cite{McLerran:1993ni,McLerran:1993ka,McLerran:1994vd}, provides a description of the parton content of a nucleus in the infinite momentum frame (IMF), where one sees it moving in the positive $x^3$ direction with a very large light cone momentum $P^+\!\gg\!\Lambda_{\text{QCD}}$. The IMF naturally gives rise to two groups of partons with vastly different dynamical features: the ``hard'' modes, which carry a large momentum fraction $p^+\!=\!xP^+$ and which could be thought of as valencelike degrees of freedom; and the ``soft'' modes, which would correspond to the small-$x$ gluons.\footnote{The separation between both groups is performed at an arbitrary momentum $\Lambda^+$. The evolution of the theory with $\Lambda^+$ is given by the Balitsky Jalilian-Marian Iancu McLerran Weigert Leonidov Kovner (B-JIMWLK) renormalization equation, which completes the CGC framework~\cite{JalilianMarian:1997gr,JalilianMarian:1997dw,Weigert:2000gi,Iancu:2000hn,Ferreiro:2001qy,Balitsky:1995ub,Kovchegov:1999yj}.}
Due to Lorentz time dilation, the latter perceive the former to be ``frozen'' in light cone time $x^+$. Also, the uncertainty principle tells us that in the IMF the hard modes appear to be sharply localized around the light cone (within a distance $\Delta x^-\!\sim\!1/p^+$). These kinematic considerations motivate the MV model to represent the valence quarks as static color currents,
\begin{align}
J^{\mu}=\delta^{\mu+}\rho(x_{\perp},x^-),
\end{align}
where the color source densities $\rho$ are assumed to be very close to a delta function in $x^-$. As for the soft modes, at high energies they are characterized by large occupation numbers (of the order of $1/\alpha_s$ for modes with transverse momenta under the saturation scale $Q_s$), which make them amenable to a description in terms of classical fields.
The dynamics of this system are encoded in the classical Yang-Mills equations
\begin{align}
[D_{\mu},F^{\mu\nu}]=J^{\nu},
\end{align}
where the valence quarks enter as the source of the gluon fields. Solving these equations in the light cone gauge we obtain the fields carried by the nucleus as pure gauge fields,
\begin{align}
\alpha^{i}(x_{\perp})=-\frac{1}{ig}U(x_{\perp})\partial^{i}U^{\dagger}(x_{\perp}).\label{gfield}
\end{align}
Here $U(x_{\perp})$ are the Wilson lines, SU($N_c$) elements defined as path-ordered exponentials,
\begin{align}\label{eq:wline}
U(x_{\perp})=\text{P}^{-}\! \exp{\left\{ -ig\int^{\infty}_{-\infty} dz^-\Phi(z^-,x_{\perp})\right\}},
\end{align}
which characterize the interaction of an external probe with the  gluon field. The fields $\Phi$ satisfy
\begin{align}
-\nabla^2_{\perp}\Phi(x^-,x_{\perp})=\tilde{\rho}(x^-,x_{\perp}),\label{aux}
\end{align}
where $\tilde{\rho}$ is the color charge density in the covariant gauge. The field of \eqref{gfield} represents the small-$x$ gluons carried by one nucleus moving close to the speed of light in the positive $x^3$ direction. Similar expressions apply in the case of a nucleus moving in the opposite direction (up to a $x^-\rightarrow x^+$ change).

In the case of a collision between two nuclei, first analyzed in Ref.~\cite{Kovner:1995ts}, the current that enters in the Yang-Mills equations has the following form:
\begin{align}
J^{\mu}=\delta^{\mu+}\rho_1(x_{\perp},x^-)+\delta^{\mu-}\rho_2(x_{\perp},x^+),
\end{align}
where the subscripts 1, 2 label the nuclei moving in the positive and negative directions, respectively. In order to solve these equations, one has to separately consider different regions in spacetime. In the forward light cone ($x^{\pm}\!>\!0$), one can parametrize the solutions in the form
\begin{align}
A^{i}(\tau,x_{\perp})&\,=\alpha^{i}(\tau,x_{\perp}),\label{ppp1}\\
A^{\pm}(\tau,x_{\perp})&\,=\pm x^{\pm}\alpha(\tau,x_{\perp}).\label{ppp2}
\end{align}
This form enforces the Fock-Schwinger gauge condition $x^+A^-+x^-A^+\!=\!0$, which acts as a sort of interpolation of the light cone gauges of each nucleus. Combining Eqs.\ (\ref{ppp1}) and (\ref{ppp2}) with the fields carried by the individual nuclei before the collision (i.e.\ $x^{\pm}\!<\!0$), the ansatz over the entire spacetime reads
\begin{align}
A^{\pm}\,&=\pm\theta(x^+)\theta(x^-)x^\pm \alpha(\tau,x_{\perp}),\\
A^{i}\;&=\theta(x^-)\theta(-x^+)\alpha_{1}^{i}(x_{\perp})\nonumber\\
& +\theta(x^+)\theta(-x^-)\alpha_{2}^{i}(x_{\perp})+\theta(x^+)\theta(x^-)\alpha^{i}(\tau,x_{\perp}).
\end{align}
No analytical solution has been found thus far for the system defined above at finite $\tau$. It is possible, however, to find exact expressions of the fields generated an infinitesimal proper time after the collision ($\tau\!=\!0^+$) in terms of $\alpha_{1,2}$. This is done by requiring the Yang-Mills equations to be regular at $\tau\!=\!0$, which leads to the following boundary conditions:
\begin{align}
\label{eq:kmw1}
\alpha^{i} (\tau=0^+\!,x_{\perp})&= \alpha_{1}^{i}(x_{\perp})+\alpha_{2}^{i}(x_{\perp}),\\
\label{eq:kmw2}
\alpha (\tau=0^+\!,x_{\perp})&= \frac{ig}{2} \left[ \alpha_{1}^{i}(x_{\perp}) , \alpha_{2}^{i}(x_{\perp}) \right].
\end{align}
These expressions, along with the Yang-Mills equations, constitute a system that yields a unique solution for $\tau$-evolution once the color source densities $\rho_{1,2}$ are given. 

In the MV model, the color charges are assumed to be stochastic variables that obey Gaussian statistics. All the information on the correlation functions of the color charges is contained in the two-point function, which in the MV model is taken to be
\begin{align}
\langle\rho^{a}(x_{\perp})\rho^{b}(y_{\perp})\rangle=g^2\mu^2\delta^{ab}\delta^{(2)}(x_{\perp}-y_{\perp}),\label{2pfMV}
\end{align}
where the variance of the distribution, $\mu^2$, is a parameter proportional to the color source number density and related to the saturation scale $Q_s$. This is the basic building block for the calculation of observables in the MV model, which are computed as ensemble averages over the color charge distributions.

In this paper, we will need higher point correlators of gluon fields in order to calculate the energy density two-point function.
We will calculate these in a Gaussian approximation, by which we retain the feature that every higher point correlator can be expressed in terms of a two-point function. However, we will use different functional forms of the relevant two-point function, generalizing the local correlator~\nr{2pfMV}.
The building block of our calculations is the correlator of two gluon fields in coordinate space,
\begin{align}
\langle \alpha^{a,i}(x_{\perp})\alpha^{b,j}(y_{\perp})\rangle\!=\frac{\delta^{ab}}{2}\bigg(&\!\delta^{ij}G(x_{\perp},y_{\perp})\nonumber\\
&+\!\left(\delta^{ij}-2\frac{r^{i}r^{j}}{r^2}\right)\!h(x_{\perp},y_{\perp})\!\bigg),\label{Fields2pf}
\end{align}
where $r_{\perp}\!=\!x_{\perp}\!-y_{\perp}$ and $r\!=\!|r_{\perp}|$. The functions $G(x_{\perp},y_{\perp})$, $h(x_{\perp},y_{\perp})$ are related to the unpolarized $\hat{G}(b_{\perp},k_{\perp})$ and linearly polarized $\hat{h}(b_{\perp},k_{\perp})$ gluon distributions by the following Hankel transforms \cite{Lappi:2017skr}:
\begin{align} \label{eq:hankel1}
G(x_{\perp},y_{\perp})&=\frac{1}{2\pi}\int dk k J_0(k r) \,\hat{G}\!\left(\frac{x_{\perp}\!+\!y_{\perp}}{2},k\right),\\
\label{eq:hankel2}
h(x_{\perp},y_{\perp})&=\frac{1}{2\pi}\int dk k J_2(k r) \,\hat{h}\!\left(\frac{x_{\perp}\!+\!y_{\perp}}{2},k\right).
\end{align}
As in this paper we work mainly in coordinate space, we will henceforth refer to $G(x_{\perp},y_{\perp})$ and $h(x_{\perp},y_{\perp})$ simply as the unpolarized and linearly polarized gluon distributions, respectively.  In Gaussian models, the polarized and unpolarized gluon distributions are not independent, but can both be explicitly related to the gluon dipole amplitude~\cite{Metz:2011wb,Lappi:2017skr}, which contains all the information about the strong interactions involved in deep inelastic scattering (DIS) processes.  Within the CGC framework, one can use different parametrizations of this object. We will in this work compare the Golec-Biernat Wusthoff (GBW) dipole distribution to a variant of the MV model that is regularized in a specific way, as detailed below. 

The GBW model \cite{GolecBiernat:1998js} was originally introduced as a purely phenomenological parametrization to fit DIS data. The  GBW dipole amplitude reads
\begin{align}
D_{\text{GBW}}(r_{\perp})=\exp{\left\{-\frac{N^2_c-1}{2N^2_c}\frac{Q_s^2r^2}{4}\right\}},
\end{align}
and the corresponding gluon distributions become
\begin{align}
G_{\text{GBW}}(r_{\perp})&\,=\frac{Q_s^2}{g^2N_c}\frac{1-\exp{\left\{-\frac{Q_s^2r^2}{4}\right\}}}{Q_s^2r^2/4},\label{GBW1}\\
h_{\text{GBW}}(r_{\perp})&\,=0\,.\label{GBW2}
\end{align}

The MV model dipole amplitude, on the other hand, is obtained by starting from the color charge correlator~\nr{2pfMV} that is local in coordinate space. In order to obtain the dipole amplitude and the gluon field correlator, one needs to spread out the delta function-like color charges in the longitudinal coordinate $x^-$ to properly calculate expectation values of the path ordered exponentials~\nr{eq:wline}. We denote the number density of color charges integrated over $x^-$ that appears in this procedure by $\bar{\mu}^2$. One also 
needs to regularize the long range Coulomb tails of the color field, i.e.\ to invert the Laplace operator in \eqref{aux}. In the literature, this is often done by introducing a ``gluon mass'' infrared regulator $m$ (see e.g.\  Ref.~\cite{Lappi:2007ku}). After this procedure~\cite{JalilianMarian:1996xn,Lappi:2007ku,Lappi:2017skr}, the dipole amplitude is given by
\begin{align}
D_{\text{MV}}(r_{\perp})=\exp{\left\{\frac{N^2_c-1}{2N_c}\frac{g^4\bar{\mu}^2}{4\pi m^2}\left(mrK_1(mr)-1\right)\right\}}.
\end{align}
The corresponding gluon distributions read
\begin{align}
\label{eq:mvG}
G_{\text{MV}}(r_{\perp})\!=&\,\frac{g^2\bar{\mu}^2}{4\pi N_c}\left(mrK_1(mr)-2K_0(mr)\right)
\\
\nonumber
&\times\frac{1-\exp{\left\{\frac{g^4\bar{\mu}^2N_c}{4\pi m^2}\left(mrK_1(mr)-1\right)\right\}}}{\frac{g^4\bar{\mu}^2}{4\pi m^2}\left(mrK_1(mr)-1\right)},\\
\label{eq:mvH}
h_{\text{MV}}(r_{\perp})\!=&\,-\frac{g^2\bar{\mu}^2}{4\pi N_c}mrK_1(mr)
\\ \nonumber
&\times\frac{1-\exp{\left\{\frac{g^4\bar{\mu}^2N_c}{4\pi m^2}\left(mrK_1(mr)-1\right)\right\}}}{\frac{g^4\bar{\mu}^2}{4\pi m^2}\left(mrK_1(mr)-1\right)}.
\end{align}
These expressions offer a wider range of possibilities for phenomenology, as they allow to pinpoint the contribution of linearly polarized gluons to the early fluctuations of glasma properties. 

\begin{figure}
\centering
\includegraphics[width=0.42\textwidth]{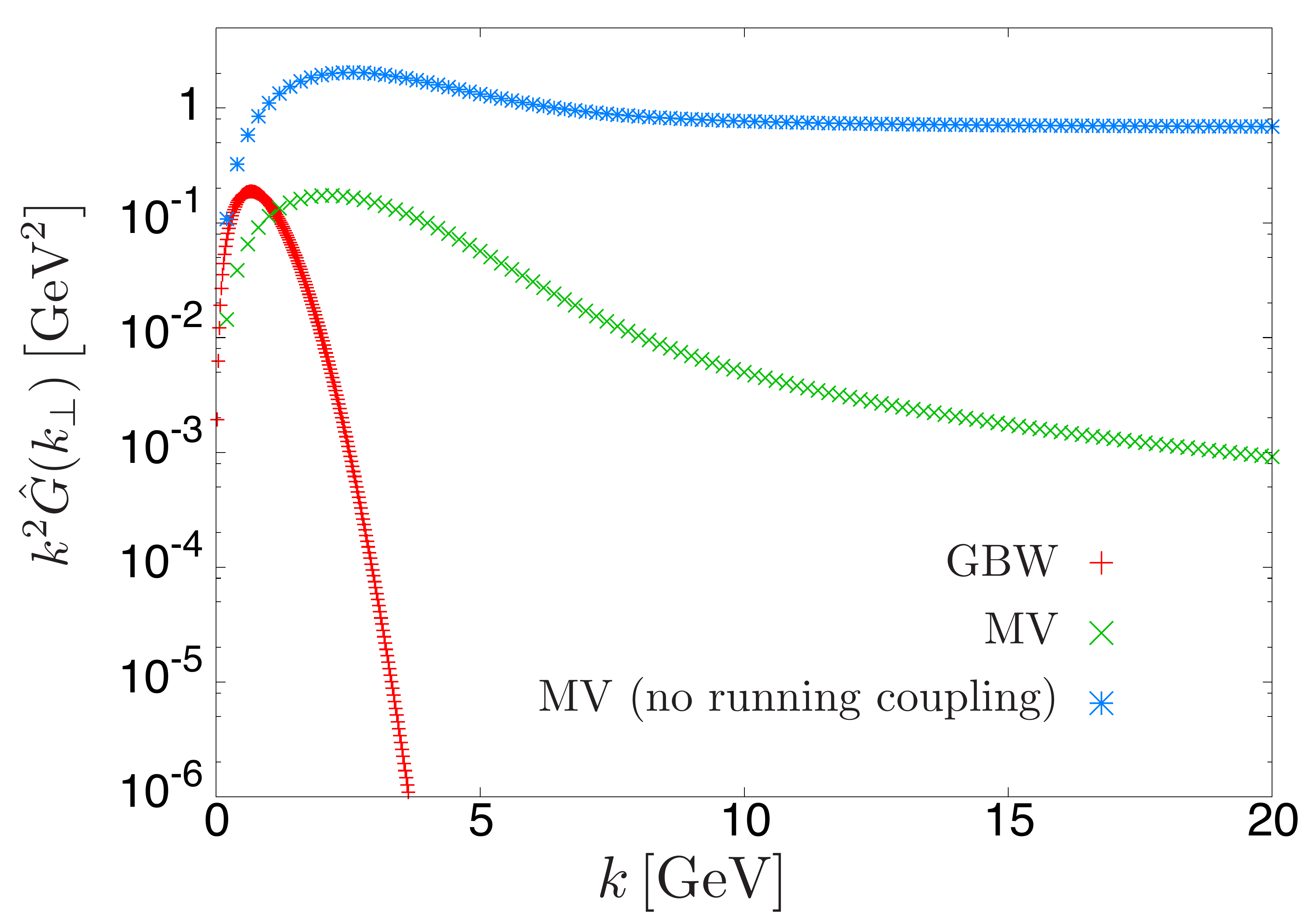}
\caption{Unpolarized gluon distributions in momentum space corresponding to different models. The distribution labeled ``MV'' is the one used here, \eqref{eq:mvG_rc}, whereas the one labeled ``MV (no running coupling)" is the one defined by \eqref{eq:mvG}.}
\label{fourier}
\end{figure} 

An essential feature of the MV model is that in the UV limit ($r\!\rightarrow\!0$) the unpolarized gluon distribution $G_{\text{MV}}$ diverges logarithmically, while the polarized one approaches a constant. In momentum space, this corresponds to the unintegrated gluon distribution in \eqref{eq:hankel1} behaving as $\sim\!1/k^2$ at large momenta. In itself, this is one of the physically attractive features of the MV model, since it makes the \emph{integrated} gluon distribution grow logarithmically with the hard scale $Q^2$, corresponding to a rudimentary approximation of the Dokshizer Gribov Lipatov Altarelli Parisi (DGLAP) evolution equations. However, it is problematic for the calculation of the $\tau\!=\!0^+$ glasma energy density, which is proportional to $G(r\!=\!0)$. As discussed in detail in \cite{Lappi:2006hq}, the glasma energy density at finite $\tau$ is finite; in the numerical classical Yang-Mills (CYM) calculations, this UV divergence at precisely $\tau\!=\!0^+$ is regularized by the lattice spacing and would in fact be divergent in the continuum limit. 
Here we will regularize this divergence by using a running coupling prescription. The choice we make is to assume that the coupling $g$ appearing as an overall coefficient (but not the coupling in the exponent) in Eqs.~\nr{eq:mvG} and~\nr{eq:mvH} runs with the distance $r$ according to
\begin{equation}
g^2(r^2)=\frac{g^2(\bar{\mu}^2)}{\ln\left(\frac{4e^{-2\gamma_e-1}}{m^2r^2}+e\right)}\label{rc1}.
\end{equation}
In the UV or small-$r$ limit, the factor on the second lines of Eqs.~\nr{eq:mvG} and~\nr{eq:mvH} approaches unity. The leading behavior of the unpolarized gluon distribution $G_{\text{MV}}(r_{\perp})$ is then dictated by the $K_0(mr)$ function. Taking the coupling in the prefactor to be the running coupling \nr{rc1} and developing the Bessel functions, the unpolarized distribution becomes
\begin{multline}
G_{\text{MV}}(r_{\perp}) \quad \underset{r\to 0}{\longrightarrow} \quad
\frac{g^2(r^2) \bar{\mu}^2}{4\pi} \ln\left(\frac{4e^{-2\gamma_e-1}}{m^2r^2}\right)
\\
= \frac{g^2(\bar{\mu}^2)\bar{\mu}^2}{4 \pi}
\frac{\ln\left(\frac{4e^{-2\gamma_e-1}}{m^2r^2}\right)}{\ln\left(\frac{4e^{-2\gamma_e-1}}{m^2r^2}+e\right)} 
\\
\underset{r\to 0}{\longrightarrow} \quad
\frac{g^2(\bar{\mu}^2)\bar{\mu}^2}{4 \pi}. 
\end{multline}
This has a nice physical interpretation: the total gluon distribution as probed by a small probe $r \to 0$ is proportional to the number density of colored particles $\bar{\mu}^2$ times a coupling evaluated at the scale corresponding to this number density $g^2(\bar{\mu}^2)$. 

In the dilute limit, the number density of color charges $\bar{\mu}^2$ is the variable that has a clear physical meaning. In the saturation regime, on the other hand, we should not be discussing in terms of the number density, but in terms of the saturation scale $Q_s$, which is defined as the inverse of the probe size $r$ for which the nonlinear behavior of the exponential on the second lines of Eqs.~\nr{eq:mvG} and~\nr{eq:mvH} starts to matter. Thus, we eliminate the explicit dependence on $\bar{\mu}^2$ in favor of the saturation scale, defined as
\begin{align}
Q^2_s= \frac{g^4(\bar{\mu}^2) }{4\pi}\bar{\mu}^2N_c.\label{rc2}
\end{align}
After this procedure, our explicit expression for the gluon distributions is
\begin{align}
\label{eq:mvG_rc}
G_{\text{MV}}(r_{\perp})\!=&\,\frac{Q_s^2}{g^2(\bar{\mu}^2)N_c}\frac{\left(mrK_1(mr)-2K_0(mr)\right)}{\ln\left(\frac{4e^{-2\gamma_e-1}}{m^2r^2}+e\right)}
\\
\nonumber
&\times\frac{1-\exp{\left\{\frac{Q_s^2}{m^2}\left(mrK_1(mr)-1\right)\right\}}}{\frac{Q_s^2}{m^2}\left(mrK_1(mr)-1\right)},\\
\label{eq:mvH_rc}
h_{\text{MV}}(r_{\perp})\!=&\,-\frac{Q_s^2}{g^2(\bar{\mu}^2)N_c}\frac{mrK_1(mr)}{\ln\left(\frac{4e^{-2\gamma_e-1}}{m^2r^2}+e\right)}
\\ \nonumber
&\times\frac{1-\exp{\left\{\frac{Q_s^2}{m^2}\left(mrK_1(mr)-1\right)\right\}}}{\frac{Q_s^2}{m^2}\left(mrK_1(mr)-1\right)}.
\end{align}
The mass regulator $m$ is parametrically a confinement scale quantity, and in this work, we will take its value as $m\!=\!0.1\,Q_s$.
With such a prescription, the gluon distributions $G_{\text{MV}}$ and $h_{\text{MV}}$ tend to the corresponding GBW model expressions in the UV limit,
\begin{align}
\lim\limits_{r\rightarrow0} G_{\text{MV}}(r_{\perp}) &= \frac{Q_s^2}{g^2N_c},\\
\lim\limits_{r\rightarrow0} h_{\text{MV}}(r_{\perp}) &= 0\,.
\end{align}

A related procedure was previously applied in \cite{Lappi:2017skr} in the calculation of correlators at $\tau\!=\!0^+$ by writing the MV model results for the gluon distributions in such a way that there is an explicit factor of $g$ in the exponent (using the fact that $Q_s\sim g^2\mu$). By taking this coupling to run, one eliminates all the logarithms of $mr$ at small $r$ from the Bessel functions and arrives at a form that in practice is similar to the GBW model.  In contrast, here we apply the running coupling prescription defined by Eqs.\ (\ref{rc1}) and (\ref{rc2}) only to the prefactors on the  first lines of Eqs.\ (\ref{eq:mvG}) and (\ref{eq:mvH}) and not to the exponent. 

The difference between the running coupling approaches in different sources in the literature amounts to some extent to the question of whether the coupling should run such that the saturation scale $Q_s$, the color charge density $g^2\bar{\mu}^2$, or the number density of color charged particles $\bar{\mu}^2$ stays fixed. Ultimately, different approaches used in the literature, including ours, are somewhat \textit{ad hoc}. They should be judged on their effect on physics, where our current approach is constructed to preserve a bit more the features of the fixed coupling MV model.  The effect of this modification is better observed in the momentum space distributions, shown in \figref{fourier}. Here we can see that our prescription indeed preserves a power law tail in the unpolarized gluon distribution. However, the power-law is visibly steeper than with the fixed coupling MV model, which distinctly yields $\lim\limits_{k\rightarrow\infty}\hat{G}(k_{\perp})\sim1/k^2$. Such a  steeper power is required for our calculation here, in order to obtain  finite results in coordinate space. 

Another difference between our approach and the one used in \cite{Lappi:2017skr} lies in the exact form of Eq.~\nr{rc1}, specifically in the $e^{-2 \gamma_E-1}$ factor contained in the argument of the logarithm in the denominator. With such a prescription, the subleading terms in the $mr\!\rightarrow\!0$ limit in Eqs.~\nr{eq:mvG_rc} and \nr{eq:mvH_rc} cancel exactly, resulting in a smoother distribution in coordinate space.

Having now discussed our approach for the gluon distribution, where Eqs.~\nr{Fields2pf}, \nr{GBW1}, \nr{GBW2}, \nr{eq:mvG_rc} and \nr{eq:mvH_rc} are the essential ingredients that we will use, let us move to what is needed to calculate the energy density two-point function. This requires one to have a four-point function of the color fields, which could be characterized as a two-gluon distribution in the colliding nuclei,  $\langle \alpha^{a}\alpha^{b}\alpha^{c}\alpha^{d}\rangle$. This is a highly complex object with a rich color structure whose full calculation is discussed in depth in \cite{Albacete:2018bbv} for the simpler case of two and three transverse positions. It could, in principle, be fully evaluated in the Gaussian approximation. Doing so would, however, be rather complicated. For the sake of simplicity, we shall thus in the present work assume that the averaging of gluon fields obeys Gaussian dynamics,
\begin{align}
\langle &\,\alpha^{i,a}_x\alpha^{j,b}_y\alpha^{k,c}_u\alpha^{l,d}_v\rangle=\langle \alpha^{i,a}_x\alpha^{j,b}_y\rangle\langle \alpha^{k,c}_u\alpha^{l,d}_v\rangle\nonumber\\
&+\langle  \alpha^{i,a}_x\alpha^{k,c}_u\rangle\langle \alpha^{j,b}_y\alpha^{l,d}_v\rangle+\langle \alpha^{i,a}_x\alpha^{l,d}_v\rangle\langle\alpha^{j,b}_y\alpha^{k,c}_u\rangle,\label{Glasmagr}
\end{align}
just like the color source densities do. Here we have adopted the shorthand notation $\alpha^{i,a}_{x}\!\equiv\!\alpha^{i,a}(x_{\perp})$. The Wick theorem-like decomposition featured in \eqref{Glasmagr} is known in the literature as the glasma graph approximation, and it has been used in many phenomenological studies of semihard two-particle correlations in the dense-dense collision regime~\cite{Dusling:2009ni,Dumitru:2010iy,Dusling:2012iga,Dusling:2012cg,Dusling:2012wy}. It will also greatly simplify the calculations presented in this paper. This assumption has been shown to yield exact results in the UV limit of \eqref{2pfgen}, where the connected, nonlinear contributions computed in Ref.~\cite{Albacete:2018bbv} vanish. In this regime, the dynamics linearize in such a way that a Gaussian distribution for the color source densities is effectively mapped onto another one for the gauge fields. Although this approximation breaks down for correlation distances larger than $1/Q_s$, it provides valuable analytical insight for the quantities computed in the present work. Our linearized approach for the time evolution would, in itself, be straightforwardly generalizable to a full nonlinear Gaussian calculation of the gauge field four-point function at $\tau=0^+$.

\subsection{Glasma correlators at $\tau\!=\!0^+$}\label{tauzero}

The structure that emerges right after a HIC in the CGC picture is characterized by the presence of purely longitudinal chromoelectric and chromomagnetic fields. In the Fock-Schwinger gauge used for the matching at $\tau=0$, these fields are given by
\begin{align}
E^{\eta}(\tau=0^+,x_{\perp})\!\equiv\!E^{\eta}_0(x_{\perp})\!=&-ig\,\delta^{ij}[\alpha^{i}_{1}(x_{\perp}),\alpha^{j}_{2}(x_{\perp})]\label{elmag1}\\
B^{\eta}(\tau=0^+,x_{\perp})\!\equiv\!B^{\eta}_0(x_{\perp})\!=&-ig\,\epsilon^{ij}[\alpha^{i}_{1}(x_{\perp}),\alpha^{j}_{2}(x_{\perp})],\label{elmag2}
\end{align}
in terms of the gluon fields carried by each nucleus prior to the collision, $\alpha^{i}_{1,2}$. As Eqs.\ (\ref{elmag1}) and (\ref{elmag2}) are the only nonvanishing components of the field strength tensor at $\tau\!=\!0^+$, the gauge invariant energy density $\varepsilon$ and the divergence of the Chern-Simons current $\dot{\nu}$ have the following simple forms:
\begin{align}
\varepsilon(\tau=0^+,x_{\perp})\!\equiv\varepsilon_0(x_{\perp})\!=&\,\text{Tr}\{E^{\eta}_0E^{\eta}_0+B^{\eta}_0B^{\eta}_0\}\label{edens},\\
\dot{\nu}(\tau=0^+,x_{\perp})\!\equiv\dot{\nu}_0(x_{\perp})\!=&\,\text{Tr}\{E^{\eta}_0B^{\eta}_0\}.
\end{align}
The expectation values of these quantities are well-known results in the CGC literature \cite{Lappi:2006hq,Lappi:2006fp}. Writing the previous expressions in terms of the gluon fields and then substituting the corresponding correlators [\eqref{Fields2pf}], we obtain~\cite{Lappi:2006hq}
\begin{align}
\langle\varepsilon_0\rangle=&\,\frac{g^2}{2}N_c(N_c^2-1)G_1(x_{\perp},x_{\perp})G_2(x_{\perp},x_{\perp})\label{edens1},\\
\langle\dot{\nu}_0\rangle=&\,0\,,
\end{align}
where $G_{1,2}$ correspond to the unpolarized gluon distributions of nuclei 1 and 2, respectively. In this result, we can see that linearly polarized gluon distributions do not contribute to these expectation values. In the GBW model, \eqref{edens1} simply reads $\langle\varepsilon_0\rangle=C_{\text{F}}\,Q^2_{s1}Q^2_{s2}/g^2$, where $C_{\text{F}}\!=\!(N_c^2-1)/(2N_c)$ is the Casimir of the fundamental representation. Note that in the MV model this object contains a logarithmic UV divergence, which we regularized by means of our running coupling prescription, but is visible in the numerical CYM calculation as a divergence in the continuum limit.

Let us now focus on the two-point functions of $\varepsilon$ and $\dot{\nu}$. Writing them in terms of the gluon fields carried by each nucleus, we obtain the following expressions:
\begin{widetext}
\begin{align}
\langle\varepsilon_0(x_{\perp})\varepsilon_0(y_{\perp})\rangle=&\,\frac{g^4}{4}(\delta^{ij}\delta^{kl}+\epsilon^{ij}\epsilon^{kl})(\delta^{i'j'}\delta^{k'l'}\!\!+\epsilon^{i'j'}\epsilon^{k'l'})\!f^{abn}f^{cdn}f^{a'b'm}f^{c'd'm}\langle\alpha^{i,a}_x\alpha^{k,c}_x\alpha^{i',a'}_y\!\alpha^{k',c'}_y\rangle_{1}\langle\alpha^{j,b}_x\alpha^{l,d}_x\alpha^{j',b'}_y\!\alpha^{l',d'}_y\rangle_{2},\label{2p0ee}\\
\langle\dot{\nu}_0(x_{\perp})\dot{\nu}_0(y_{\perp})\rangle=&\,\frac{g^4}{4}\delta^{ij}\epsilon^{kl}\delta^{i'j'}\!\epsilon^{k'l'}\!f^{abn}f^{cdn}f^{a'b'm}f^{c'd'm}\langle\alpha^{i,a}_x\alpha^{k,c}_x\alpha^{i',a'}_y\!\alpha^{k',c'}_y\rangle_{1}\langle\alpha^{j,b}_x\alpha^{l,d}_x\alpha^{j',b'}_y\!\alpha^{l',d'}_y\rangle_{2}.\label{2p0nn}
\end{align}
\end{widetext}

These objects have an intricate color algebra composition, featuring four SU($N_c$) structure constants whose indices are contracted with two four-point correlators (one per nucleus). Furthermore, the indices labeling the transverse plane vectors also contribute to the complexity of the formulas (and even more in the finite $\tau$ case). Evaluating such expressions is challenging even when employing the simplest approach available (i.e.\ adopting both the glasma graph approximation and the GBW model), which is why we used the \textit{Mathematica} package \textsc{FeynCalc} \cite{Mertig:1990an,Shtabovenko:2016sxi} to perform the algebraic manipulations required throughout the calculation process.
By applying the glasma graph approximation and adopting the GBW model, one is able to obtain the following formulas:
\begin{align}
\frac{\langle\varepsilon_0(x_{\perp})\varepsilon_0(y_{\perp})\rangle}{\langle\varepsilon_0(x_{\perp})\rangle\langle\varepsilon_0(y_{\perp})\rangle}-1=&\,\frac{3}{N_c^2-1}\Bigg[\frac{1}{3}\left(\frac{1-e^{-Q_{s}^2r^2/4}}{Q_s^2r^2/4}\right)^4\nonumber\\
&+\frac{2}{3}\left(\frac{1-e^{-Q_{s}^2r^2/4}}{Q_s^2r^2/4}\right)^2\Bigg]\label{2p0e},\\
\frac{\langle\dot{\nu}_0(x_{\perp})\dot{\nu}_0(y_{\perp})\rangle}{\langle\varepsilon_0(x_{\perp})\rangle\langle\varepsilon_0(y_{\perp})\rangle}=&\,\frac{3}{8(N_c^2-1)}\left(\frac{1-e^{-Q_{s}^2r^2/4}}{Q_s^2r^2/4}\right)^4,\label{2p0n}
\end{align}
first presented in this compact form in \cite{Lappi:2017skr}. 

By performing the full calculation of the four-point function of the fields (i.e.\ without assuming a Gaussian-like decomposition), one arrives at much lengthier expressions that, remarkably, feature different asymptotic behaviors in the $Q_{s}r\!\gg\!1$ limit ($\propto\!1/r^2$ in the case of the energy density and $\propto\!1/r^4$ for the divergence of the Chern-Simons current) \cite{Albacete:2018bbv}. We expect that, since  the considered quantities are largely UV dominated, the breakdown of the glasma graph approximation should not have a big impact on the qualitative conclusions of the present work.
Conversely, the discrepancy at large correlation distances will likely be much more relevant when considering quantities that are sensitive to the infrared. An important example of this is given by the integral of \eqref{2p0e} over the transverse plane, which is a key quantity in the calculation of eccentricity fluctuations. In fact, the infrared sensitivity enters due to the $1/r^2$ falloff featured in the full calculation. This asymptotic behavior gives rise to a correlation length that depends on the infrared scale $1/m$ through a logarithmic factor $\ln\left( Q_s/m\right)$. The study of the $\tau$ dependence of this property is left for future work.

\section{Proper time evolution}\label{evol}

Our goal in this section is to generalize the previous correlators to finite proper times, $\tau\!>\!0$. As the respective supports of the color charge sources are Lorentz contracted almost to delta functions ($\rho_{1,2}\!\sim\!\delta(x^{\pm})$), in the spacetime region $\tau>0$ the Yang-Mills equations become homogeneous, $[D_{\mu},F^{\mu\nu}]=0$. The separate components of these equations in the comoving coordinate system $(\tau,\eta,i)$ read
\begin{align}
ig\tau\left[ \alpha, \partial_{\tau}\alpha\right]-\frac{1}{\tau}\left[ D_{i},\partial_{\tau}\alpha^{i} \right]&=0,\\
\frac{1}{\tau}  \partial_{\tau}\frac{1}{\tau}\partial_{\tau}(\tau^2 \alpha) -\left[ D_{i},\left[ D^{i},\alpha \right]\right]&=0,\\
\frac{1}{\tau}  \partial_{\tau}(\tau \partial_{\tau}\alpha^{i}) -ig\tau^2\left[ \alpha,\left[ D^{j},\alpha \right]\right]-\left[ D^{j},F^{j i}\right]&=0.\label{comov3}
\end{align}
We now want to solve these equations by linearizing them, i.e.\ neglecting all the terms that are higher order in the gauge potentials $\alpha,\alpha^i$. This linearization unavoidably introduces a gauge dependence in the calculation, since any gauge transformation changes the value of the gauge potentials, and thus the value of the neglected higher order terms. However, there are several physical constraints that allow one to choose the proper gauge. First, we are solving the equations of motion in a spacetime region without any external charges that would introduce a static Coulomb field. It is therefore natural to maintain the Fock-Schwinger gauge condition $A_\tau=0$, which is achieved by refraining from $\tau$-dependent gauge transformations. Second, our physical situation is boost invariant, and it is natural to maintain this boost invariance at the level of the gauge potentials. This means that we do not allow our gauge transformation to depend on the spacetime rapidity $\eta$. These conditions restrict us to performing gauge transformations that only depend on the transverse coordinate. The general such transformation can be written as
\begin{align}
\alpha(\tau,x_{\perp})&=U(x_{\perp})\beta(\tau,x_{\perp}) U^{\dagger}(x_{\perp}),\\
\alpha^{i}(\tau,x_{\perp})&=U(x_{\perp})\left(\beta^{i}(\tau,x_{\perp})-\frac{1}{ig}\partial^{i}\right)U^{\dagger}(x_{\perp}).\label{gueich}
\end{align}
The linearized equations of motion for the transformed gauge potential $\beta$ read
\begin{align}
\partial_{\tau}\partial_{i}\beta^{i}&=0\label{eer0},\\
\frac{1}{\tau}  \partial_{\tau}\frac{1}{\tau}\partial_{\tau}(\tau^2 \beta) -\partial_{i}\partial^{i}\beta&=0\label{eer1},\\
\frac{1}{\tau}  \partial_{\tau}(\tau \partial_{\tau}\beta^{i}) -(\partial^{k}\partial_{k}\delta^{ij}-\partial^{i}\partial^{j})\beta^{j}&=0.\label{eer2}
\end{align}
A natural choice for fixing the last degree of freedom in the gauge transformation is to choose the transverse Coulomb gauge defined by the condition $\partial_i\beta^{i}=0$. This choice minimizes the amplitudes of the transverse components of the fields, and one could thus argue that it is the choice where a linearized approximation works best. Since we refrain from time-dependent gauge transformations, we can in general only enforce the Coulomb condition at one specific value of $\tau$. With the general nonlinear equations of motion the condition $\partial_i\beta^{i}=0$ is not conserved in $\tau$~\cite{Boguslavski:2018beu}. However, for the \emph{linearized} equations, the Coulomb condition is explicitly conserved in time \nr{eer0}. Thus, with linearized evolution, we can work completely in Coulomb gauge at all values of $\tau$, and this is what we shall do in the following. In this case, the equations of motion reduce to 
\begin{align}
\partial_{\tau}\frac{1}{\tau}\partial_{\tau}(\tau^2 \beta) &=\tau\partial_{i}\partial^{i}\beta\label{eer11_v2},\\
\partial_{\tau}(\tau \partial_{\tau}\beta^{i}) &=\tau\partial^{k}\partial_{k}\beta^{i} \label{eer22_v2},
\end{align}
which can be straightforwardly solved by Fourier transforming. The general plane wave solutions for the time dependence of the momentum modes are 
\begin{align}
\beta(\tau,k_{\perp}) &=\beta_0(k_{\perp}) \frac{2 J_1(\omega\tau)}{\omega\tau}\nonumber,\\
\beta^i(\tau,k_{\perp}) &=\beta^i_0(k_{\perp})J_0(\omega\tau),\label{sollsS_v2}
\end{align}
where we assume the free dispersion relation $\omega(k_{\perp})\!=\!|k_{\perp}|$.

We must now relate the initial conditions of the gauge potentials, $\beta_0(k_{\perp})$ and $\beta^i_0(k_{\perp})$, to the gauge fields of the incoming nuclei. In principle, we would do this by taking the initial conditions for the gauge fields from Eqs.~\nr{eq:kmw1} and~\nr{eq:kmw2}, and performing the gauge transformation to Coulomb gauge. This procedure was done numerically in Ref.~\cite{Blaizot:2010kh}. In this study it was found that a linear time evolution starting from such a Coulomb gauge initial condition is a very good approximation of the full CYM evolution, apart from the very soft modes whose evolution is genuinely nonlinear due to screening. Finding the required gauge transformation \nr{gueich} to Coulomb gauge is also possible analytically in the dilute-dense case~\cite{Dumitru:2001ux,Lappi:2017skr}, leading to nice $k_\perp$-factorized expressions for the gluon spectrum in proton-nucleus collisions that are often used in  phenomenology~\cite{Kovchegov:2001sc,Blaizot:2004wu} (a calculation recently extended one order higher in the color charge density in the proton~\cite{Li:2021zmf}). However, in the full dense-dense case that we want to address here, finding an explicit form for the gauge transformation $U(x_\perp)$ from \eqref{gueich} is not possible analytically. A way out of this conundrum can be found by fully exploiting the fact that we have resolved to use a linearized approximation for the time evolution in the region $\tau>0$, and that the physics we are interested in really depend on the electric and magnetic fields, not the gauge potentials themselves. Thus, in fact, we do not need an explicit form of the gauge transformation $U(x_\perp)$. Instead, it is enough to find a gauge potential that does the following:
\begin{enumerate}
    \item Satisfies the Coulomb gauge linearized equation of motion, i.e.\ is of the form~\nr{sollsS_v2}
    \item Reproduces the initial field strength \nr{elmag1} and \nr{elmag2} at $\tau\!=\!0^+$ when calculating the electric and magnetic strength from the gauge potential using the linearized approximation, consistently with the linear time evolution
\end{enumerate}
Thus, instead of finding an explicit form for the gauge transformation, we obtain the gauge potential at $\tau\!>\!0$ by performing a matching of the electric and magnetic fields to the values at $\tau\!=\!0^+$. Alternatively, one could formulate our approach in terms of linear equations of motion for the electric and magnetic fields themselves, as the Abelian Maxwell equations are usually written. This equation  can then be matched to an initial condition that includes the full nonlinear dependence in the color charges of the colliding projectiles. This procedure enables us to include the full nonlinear structure of gluon saturation in both colliding projectiles at the initial time, but nevertheless obtain an analytical expression for time evolution.

Taking this approach, it is now straightforward to fix the initial conditions in the general solution~\nr{sollsS_v2} to obtain
\begin{align}
\beta(\tau,k_{\perp}) &=\frac{\tau}{k}E^{\eta}_0(k_{\perp})J_1(k\tau)\nonumber,\\
\beta^i(\tau,k_{\perp}) &=-i\frac{\epsilon^{ij}k^j}{k^2}B^{\eta}_0(k_{\perp})J_0(k\tau)\label{solls2},
\end{align}
with $E^{\eta}_0(k_{\perp})$ and $B^{\eta}_0(k_{\perp})$ given by the Fourier transforms of Eqs.~\nr{elmag1} and~\nr{elmag2}.
The evolution is thus described simply as the propagation of the initial conditions as a free plane wave. This is a reasonable approximation for high momentum modes only and therefore appropriate for largely UV-dominated quantities such as the energy density or the divergence of the Chern-Simons current.
Correspondingly, the full electric and magnetic fields as functions of $\tau$ are given by 
\begin{align}
E^{\eta}(\tau,k_{\perp})=&\,E^{\eta}_0(k_{\perp})J_0(k\tau),\label{etau}\\
E^{i}(\tau,k_{\perp})=&\,-i\epsilon^{ij}\frac{k^{j}}{k}B^{\eta}_0(k_{\perp})J_1(k\tau),\\
B^{\eta}(\tau,k_{\perp})=&\,B^{\eta}_0(k_{\perp})J_0(k\tau),\\
B^{i}(\tau,k_{\perp})=&\,-i\epsilon^{ij}\frac{k^{j}}{k}E^{\eta}_0(k_{\perp})J_1(k\tau).
\end{align}

In order to infer the $\tau$ dependence in coordinate space, we rewrite the initial condition \eqref{elmag1} as the following Fourier transform:
\begin{align}
E^{\eta}_0(x_{\perp})\!&=\!-ig\delta^{ij}\!\!\int \!\frac{d^2k_{\perp}}{(2\pi)^2}\!\!\int \!d^2u_{\perp}[\alpha^{i}_{1}(u_{\perp}),\alpha^{j}_{2}(u_{\perp})]e^{ik_{\perp}(x-u)_{\perp}}\nonumber\\
&\equiv\!\int \frac{d^2k_{\perp}}{(2\pi)^2}E^{\eta}_0(k_{\perp})e^{ik_{\perp}x_{\perp}},
\end{align}
and thus, from \eqref{etau}, we get to
\begin{align}
E^{\eta}(\tau,x_{\perp})\!=\!-ig\delta^{ij}\!\!\int \!\frac{d^2k_{\perp}}{(2\pi)^2}\!\!\int &\!d^2u_{\perp}[\alpha^{i}_{1}(u_{\perp}),\alpha^{j}_{2}(u_{\perp})]\nonumber\\
&\times\!J_0(k\tau)e^{ik_{\perp}(x-u)_{\perp}}.\label{el_ft}
\end{align}
Similar transformations of the initial conditions are obtained for $E^{i}$ and the chromomagnetic fields.

This approach to time evolution can also be applied in the case where one of the sources is much weaker than the other. However, we observe that by taking the dilute-dense limit of the electric field,~\eqref{el_ft}, or the resulting energy density for $\tau\!>\!0$, one does not recover the known result in the dilute-dense regime. In this limit, the object representing the gluon content of the dense nucleus is a dipole distribution, whereas our calculation (as will be shown later) implies the unpolarized Weizs\"acker-Williams distributions of each nucleus. Both approaches yield the same result at $\tau\!=\!0$, since the $r_\perp=0$ limit of both gluon distributions is the same. However, the convolution with the Bessel functions introduced in \eqref{el_ft} induces their Fourier transforms to differ at $\tau\!>\!0$, because the distributions are different as functions of $p_\perp$.
This problem is not really associated with our approach to the time dependence per se, but rather to our inability to analytically Fourier transform the initial fields \eqref{elmag1} and ~\eqref{elmag2} to Coulomb gauge. In the dilute-dense limit, this gauge transformation can be done, and after this transformation our linearized approach to the time dependence would reduce to the conventional result.

Let us briefly discuss the relation of our approach to that of Refs.~\cite{Fries:2006pv,Fujii:2008km,Chen:2015wia,Carrington:2020ssh}, where one performs a systematic power expansion of the fields in orders of $\tau$. In this approach, the Yang-Mills equations are reformulated as an infinite system of equations for the coefficients of the series. Taking the initial conditions Eqs.\ (\ref{elmag1}) and (\ref{elmag2}) as the lowest order $\tau^0$, the remaining coefficients are obtained recursively. The main drawback of this approach comes in the form of UV divergences, which are carried by each term of the series. These singularities originate in the spatial derivatives and give rise to the dominant terms of the expansion.
The interpretation of this UV divergence in terms of the Bessel functions $J_{0,1}(k\tau)$ that give the (free) time dependence of the momentum modes is evident. Dimensionally, every power of $\tau$ is associated with an additional power of momentum that is integrated over, making the higher coefficients ever more UV divergent. Resumming the whole series into a Bessel function, on the other hand, actually improves the UV behavior at any nonzero value of $\tau$, because of the suppression $J_{(0,1)}(k\tau) \sim 1/\sqrt{k}$ for $k \gg 1/\tau$. This property is what makes the energy density in the MV model (even without any running coupling) UV finite at $\tau>0$, even if it is UV divergent at $\tau\!=\!0$~\footnote{For a finite UV cutoff $\Lambda$, the MV model energy density has finite limit $\Lambda \to \infty$ at $\tau>0$, but diverges as $\ln^2\Lambda$ at $\tau=0$. In the lattice calculation, the UV cutoff $\Lambda$ corresponds to $1/a$, where $a$ is the lattice spacing. Thus, one refers to the limit $\Lambda\to \infty$ as the ``continuum limit.'' From this behavior, one can deduce that if one takes the limit $\Lambda\to\infty$ first and then approaches $\tau=0^+$ from above, the continuum limit energy density will diverge as $\ln^2 1/\tau$. Thus, in the MV model at fixed coupling, a power series expansion in $\tau$ is a series expansion around a point where the continuum limit energy density is nonanalytic~\cite{Lappi:2006hq}. }. The same asymptotic behavior $J_{(0,1)}(k\tau) \sim 1/\sqrt{\tau}$ for $\tau \gg 1/Q_s \sim 1/k$ directly results in the large time $\sim 1/\tau$ behavior of the energy density. This is naturally interpreted as the effect of the boost invariant expansion of the whole system and would be difficult to reach in a power series expansion in $\tau$.
This is also pointed out in Ref.~\cite{Fujii:2008km}, where the highest-derivative terms of the Yang-Mills equations are resummed into a free field evolution. This ansatz is technically identical to considering the linearized evolution equations we use. Therefore, \eqref{solls2} can be understood as a resummation of the high momentum modes of a $\tau$ expansion of the forward light cone fields.

\subsection{One-point functions}
\label{sec:onept}

Using the formulas presented above, the $\tau$-dependent expectation values of the energy density and the divergence of the Chern-Simons current read
\begin{widetext}
\begin{align}
\langle\varepsilon(\tau,x_{\perp})\rangle\!=\!\langle\text{Tr}\{E^{\eta}E^{\eta}\!+\!B^{\eta}B^{\eta}\!+\!E^iE^i\!+\!B^iB^i\} \rangle\!=&\,\frac{g^2}{2}(\delta^{ij}\delta^{kl}\!+\epsilon^{ij}\epsilon^{kl})f^{abn}f^{cdn}\!\!\!\!\int\limits_{p,k}\int\limits_{u,v}\langle\alpha^{i,a}_u\alpha^{k,c}_v\rangle_1\langle\alpha^{j,b}_u\alpha^{l,d}_v\rangle_2\nonumber\\
&\times\left(J_0(p\tau)J_0(k\tau)-\frac{p_{\perp}\!\cdot\!k_{\perp}}{p\,k}J_1(p\tau)J_1(k\tau)\right)e^{ip_{\perp}(x-u)_{\perp}}e^{ik_{\perp}(x-v)_{\perp}},\label{tauEd}
\end{align}
\vspace{-0.75cm}
\begin{align}
\langle\dot{\nu}(\tau,x_{\perp})\rangle\!=\!\langle\text{Tr}\{E^{\eta}B^{\eta}\!+\!E^iB^i\} \rangle\!=&\,\frac{g^2}{2}f^{abn}f^{cdn}\!\!\!\!\int\limits_{p,k} \int\limits_{u,v}\langle\alpha^{i,a}_u\alpha^{k,c}_v\rangle_1\langle\alpha^{j,b}_u\alpha^{l,d}_v\rangle_2\nonumber\\
&\times\left(\delta^{ij}\epsilon^{kl}J_0(p\tau)J_0(k\tau)-\epsilon^{ij}\delta^{kl}\,\frac{p_{\perp}\!\cdot\!k_{\perp}}{p\,k}J_1(p\tau)J_1(k\tau)\right)e^{ip_{\perp}(x-u)_{\perp}}e^{ik_{\perp}(x-v)_{\perp}},\label{taunud}
\end{align}
\end{widetext}
where $\int_{p}$ stands for the integration over momentum $\int\frac{d^2p_{\perp}}{(2\pi)^2}$, while $\int_{u}$ corresponds to $\int d^2u_{\perp}$. There are some operations that can be performed on these general expressions before adopting a specific dipole model. For instance, by substituting the general two-point function~\nr{Fields2pf} in \eqref{taunud}, one directly obtains that $\langle\dot{\nu}\rangle\!=\!0$ for any value of $\tau$. This is a trivial result that reflects that our field averages are performed over a \textit{CP}-even ensemble.
Let us now focus on the more interesting case of the average energy density. This object contains Fourier transforms of Bessel functions of zeroth and first order, which can be computed analytically. Let us show this calculation explicitly. Integrating over the angular variables of the momenta $\theta_p$ and $\theta_k$, the second line of \eqref{tauEd} becomes
\begin{multline}
\int\frac{dp\,p}{(2\pi)}\frac{dk\,k}{(2\pi)}\bigg(J_0(|x-u|p)J_0(|x-v|k)J_0(p\tau)J_0(k\tau)
\\
+\cos{(\theta_{x-u}\!-\theta_{x-v})}J_1(|x\!-\!u|p)J_1(|x\!-\!v|k)J_1(p\tau)J_1(k\tau)\!\bigg).
\end{multline}
Then, applying the orthogonality condition of the Bessel functions,
\begin{align}
\int^{\infty}_{0}J_{\nu}(kr)J_{\nu}(sr)rdr=\frac{\delta(k-s)}{s},
\end{align}
we obtain
\begin{align}
\hspace{-0.2cm}\langle\varepsilon(\tau,x_{\perp})\rangle\!=\!\frac{g^2}{2}(\delta^{ij}\delta^{kl}\!+\epsilon^{ij}\epsilon^{kl})f^{abn}f^{cdn}\!\!\!\int\limits_{u,v}\!\!\langle\alpha^{i,a}_u\alpha^{k,c}_v\rangle_1 \langle\alpha^{j,b}_u\alpha^{l,d}_v\rangle_2\nonumber\\
\times\!\frac{\delta(|x_{\perp}-u_{\perp}|\!-\!\tau)}{2\pi\tau}\frac{\delta(|x_{\perp}-v_{\perp}|\!-\!\tau)}{2\pi\tau}(1+\cos(\theta_{x-u}\!-\theta_{x-v})).\label{tauEd2}
\end{align}

\begin{figure}
\includegraphics[width=0.32\textwidth]{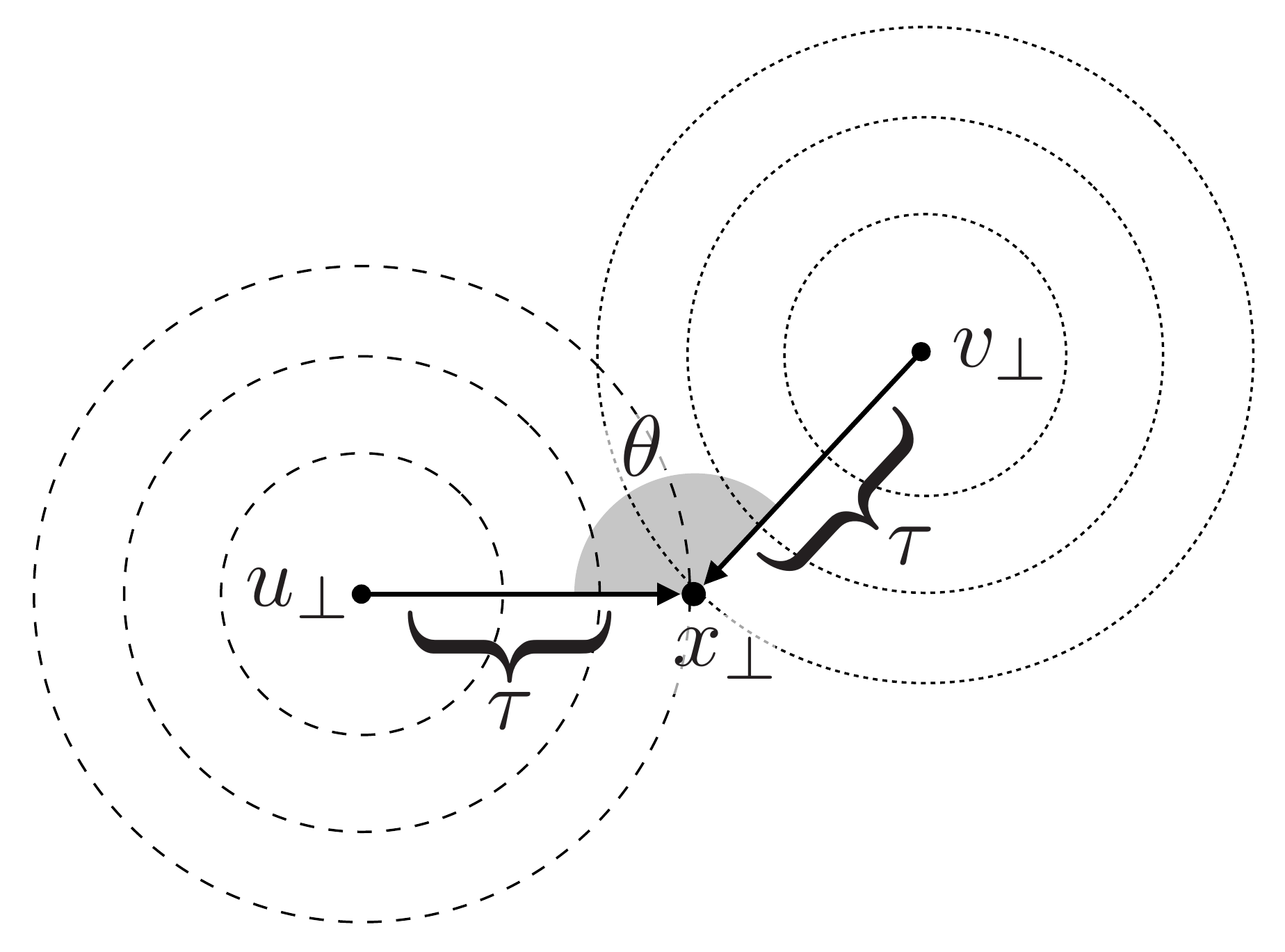}
\caption{Representation of an infinitesimal contribution to the average energy density deposited on a point $x_{\perp}$ at a proper time $\tau$. The shaded angle corresponds to the angle that controls the interference effect, $\theta\!=\!\theta_{x-u}\!-\theta_{x-v}$.}
\label{wafs}
\end{figure}

The average energy density deposited on a transverse point $x_{\perp}$ at a certain proper time $\tau$ results from the interference of the Weizs\"acker-Williams distributions characterizing each nucleus, $\langle\alpha^{i,a}_u\alpha^{k,c}_v\rangle$. In Gaussian models, it is possible to interpret this object as the gluon distribution of a nucleus probed by a gluon dipole with ``legs" on transverse positions $u_{\perp}$ and $v_{\perp}$. At $\tau\!=\!0^+$, the only point that contributes to the average energy density at $x_{\perp}$ is $x_{\perp}$, and thus this quantity is proportional to the zero-sized dipole $\langle\alpha^{i,a}_x\alpha^{k,c}_x\rangle$ (a divergent object). At finite values of $\tau$, however, the average energy density deposited at $x_{\perp}$ has received contributions from all those dipoles whose legs sit at a distance $\tau$ from $x_{\perp}$. As we have approximated the $\tau$ evolution with linear free field equations, these contributions evolve as circular wave fronts that propagate on the transverse plane at the speed of light, thus converging on the point $x_{\perp}$ at a proper time $\tau$. Moreover, the $\tau^{-1}$ factors account for the attenuation caused by this expansion. The result of the interference is controlled by the angle between the vectors that link $u_{\perp}$ and $v_{\perp}$ to $x_{\perp}$ (see \figref{wafs}). There are thus integration regions where the interference is destructive, making no contribution to the average energy density at that point. This happens when the considered point is located in the middle of the gluon dipole, with $\cos(\theta_{x-u}\!-\theta_{x-v})\!=\!-1$. On the other hand, the radiation emanating from point $u_\perp\!=\!v_\perp$ to $x_\perp$ interferes constructively.

\begin{figure*}
\centering
\includegraphics[width=0.46\textwidth]{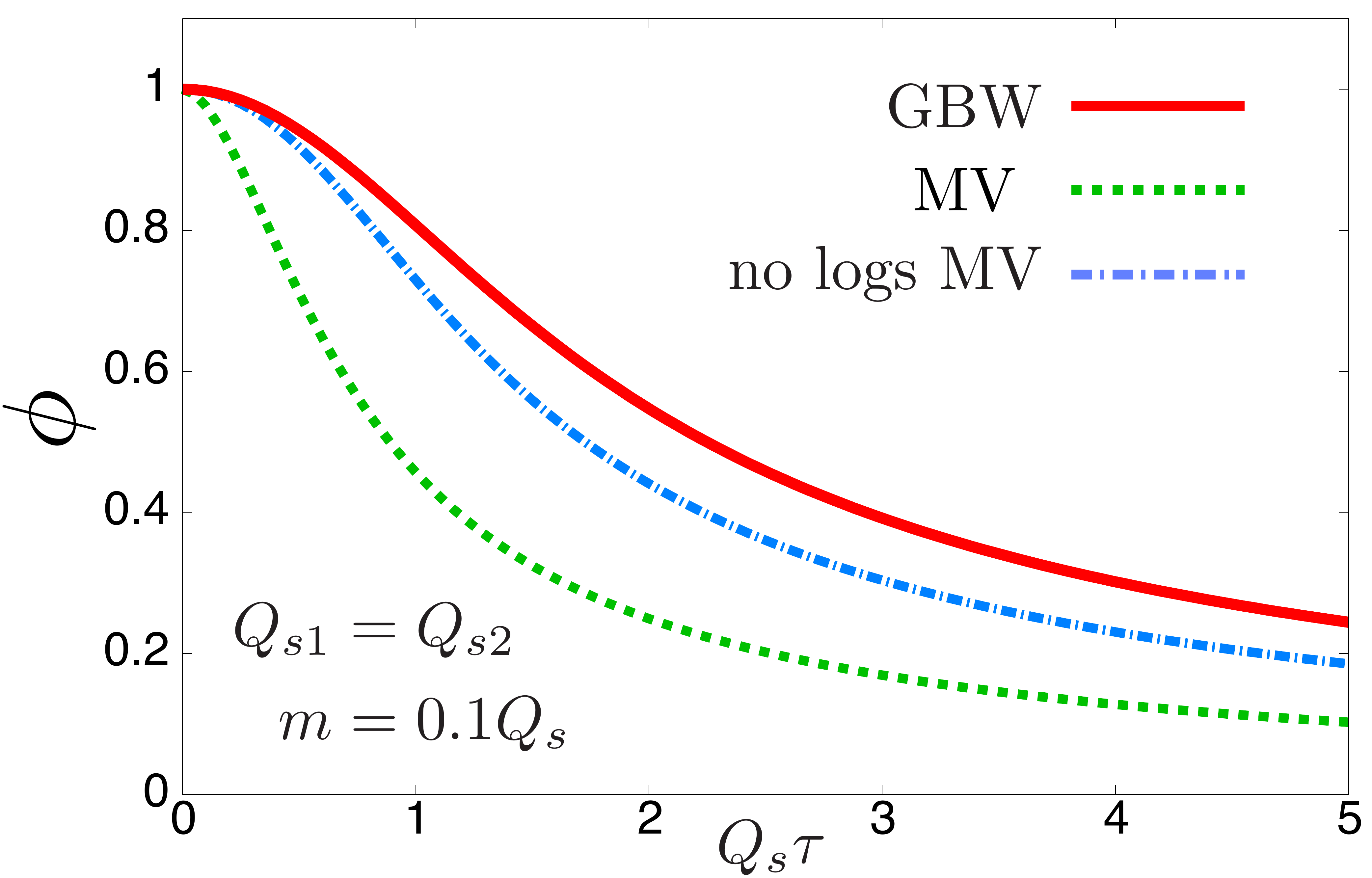}
\caption{Energy density dilution function for the GBW (thick curve) and MV models. Here we distinguish between the results provided by the running coupling prescription applied in this work (dashed curve) and the one adopted in \cite{Lappi:2017skr} (``no logs MV''). Note that the MV model without any running coupling or UV cutoff would have an infinite $\left<\varepsilon_0\right>$ and a finite  $\langle\varepsilon(\tau)\rangle$ for $\tau>0$; thus, plotting the ratio $\phi=\langle\varepsilon(\tau)\rangle/\left<\varepsilon_0\right>$ would not make sense. }
\label{Result0b}
\end{figure*}

We can compute almost all of the integrals of \eqref{tauEd2} analytically by adopting the GBW model. First, we perform a change of variables from ($u_{\perp}$, $v_{\perp}$) to ($s_{\perp}\!=\!x_{\perp}-u_{\perp}$, $t_{\perp}\!=x_{\perp}-v_{\perp}$). Then, we can integrate over $|s_{\perp}|$ and $|t_{\perp}|$, obtaining
\begin{multline}
\langle\varepsilon(\tau)\rangle\!=\!\frac{g^2}{2}N_c(N_c^2-1)\!\!\!\int^{2\pi}_{0}\frac{d\theta_s}{2\pi}\frac{d\theta_t}{2\pi}(1+\cos(\theta_{s}\!-\!\theta_{t}))
\\
\times\!\!\left(\frac{Q_{s1}^2}{g^2N_c}\frac{1-\exp{\!\left\{-\frac{Q_{s1}^2\tau^2}{2}(1-\cos(\theta_{s}\!-\!\theta_{t}))\right\}}}{\frac{Q_{s1}^2\tau^2}{2}(1-\cos(\theta_{s}\!-\!\theta_{t}))}\right)\times\left(1\!\rightarrow2\right).\label{tauEd3b}
\end{multline}
Now, after performing one more variable change to $\Theta\!=\!\theta_s\!-\theta_t$, we can integrate over the other angle and finally arrive at
\begin{multline}
\langle\varepsilon(\tau)\rangle\!=\!\frac{g^2}{2}N_c(N_c^2-1)\!\!\!\int^{2\pi}_{0}\frac{d\Theta}{2\pi}(1+\cos(\Theta))
\\
\times\!\!\left(\frac{Q_{s1}^2}{g^2N_c}\frac{1-\exp{\!\left\{-\frac{Q_{s1}^2\tau^2}{2}(1-\cos(\Theta))\right\}}}{\frac{Q_{s1}^2\tau^2}{2}(1-\cos(\Theta))}\right)\!\!\!\times\!\!\left(1\!\rightarrow2\right)
\\
\equiv\langle\epsilon_0\rangle\!\times\phi(Q_{s1}\tau,Q_{s2}\tau).\label{tauEd3}
\end{multline}
This result consists of the initial expectation value for the energy density $\langle\epsilon_0\rangle$ multiplied by a nontrivial dilution factor $\phi$ which depends on the dimensionless products $Q_{s1,2}\tau$. Increasing the value of the saturation scale is equivalent to fast-forwarding $\tau$ evolution, as this provokes dilution to occur faster. One can also see that at large $\tau$ only angles $\Theta \lesssim 1/(Q_s \tau)$ contribute to the integral, thus leading to the time dependence $\langle\varepsilon(\tau)\rangle \sim 1/\tau$.
In \figref{Result0b}, we display $\phi$ as a function of $Q_s\tau$ for both GBW and MV models. Here one can see that at $\tau\!\sim\!1/Q_s$ our MV model prescription (dashed curve) yields almost twice as much energy density dilution as the GBW model (thick curve). This is interpreted as the effect of the larger power-law tail of fast-evolving high $k_\perp$ modes in the MV model. Also note that, if we apply the running coupling correction defined by \eqref{rc1} to suppress all the logarithms introduced by the MV model (dot-dashed curve, corresponding to the prescription used in Ref.~\cite{Lappi:2017skr}), the difference with the GBW model at this point is much less prominent.

\subsection{Two-point functions}
Let us now focus on the calculation of two-point functions. By expanding the $\tau$-dependent chromoelectric and chromomagnetic fields~\nr{etau} in terms of the initial conditions, we obtain the following expressions:
\begin{widetext}
\begin{align}
\langle\varepsilon(\tau,x_{\perp})\varepsilon(\tau,y_{\perp})\rangle\!=&\,\frac{g^4}{4}(\delta^{ij}\delta^{kl}\!+\epsilon^{ij}\epsilon^{kl})(\delta^{i'\!j'\!}\delta^{k'\!l'\!}\!+\epsilon^{i'\!j'\!}\epsilon^{k'\!l'\!})f^{abn}f^{cdn}f^{a'\!b'\!m}f^{c'\!d'\!m}\!\!\!\!\int\limits_{p,k}\int\limits_{\bar{p},\bar{k}}\int\limits_{u,v}\int\limits_{\bar{u},\bar{v}}\langle\alpha^{i,a}_u\alpha^{k,c}_{\bar{u}}\alpha^{i'\!\!,a'\!\!}_v\alpha^{k'\!\!,c'\!\!}_{\bar{v}}\rangle\langle\alpha^{j,b}_u\alpha^{l,d}_{\bar{u}}\alpha^{j'\!\!,b'\!\!}_v\alpha^{l'\!\!,d'\!\!}_{\bar{v}}\rangle\nonumber\\
&\times\!\!\left(J_0(p\tau)J_0(\bar{p}\tau)\!-\frac{p_{\perp}\!\cdot\!\bar{p}_{\perp}}{p\,\bar{p}}J_1(p\tau)J_1(\bar{p}\tau)\right)\!\!\left(J_0(k\tau)J_0(\bar{k}\tau)\!-\frac{k_{\perp}\!\cdot\!\bar{k}_{\perp}}{k\,\bar{k}}J_1(k\tau)J_1(\bar{k}\tau)\right)\nonumber\\
&\times\!e^{ip_{\perp}\!(x-u)_{\perp}}e^{ik_{\perp}\!(y-v)_{\perp}}e^{i\bar{p}_{\perp}\!(x-\bar{u})_{\perp}}e^{i\bar{k}_{\perp}\!(y-\bar{v})_{\perp}}\nonumber\\
=&\,\frac{g^4}{4}(\delta^{ij}\delta^{kl}\!+\epsilon^{ij}\epsilon^{kl})(\delta^{i'\!j'\!}\delta^{k'\!l'\!}\!+\epsilon^{i'\!j'\!}\epsilon^{k'\!l'\!})f^{abn}f^{cdn}f^{a'\!b'\!m}f^{c'\!d'\!m}\!\!\!\int\limits_{u,v}\int\limits_{\bar{u},\bar{v}}\langle\alpha^{i,a}_u\alpha^{k,c}_{\bar{u}}\alpha^{i'\!\!,a'\!\!}_v\alpha^{k'\!\!,c'\!\!}_{\bar{v}}\rangle\langle\alpha^{j,b}_u\alpha^{l,d}_{\bar{u}}\alpha^{j'\!\!,b'\!\!}_v\alpha^{l'\!\!,d'\!\!}_{\bar{v}}\rangle\nonumber\\
&\times\!\frac{\delta(|x_{\perp}-u_{\perp}|\!-\!\tau)}{2\pi\tau}\frac{\delta(|x_{\perp}-\bar{u}_{\perp}|\!-\!\tau)}{2\pi\tau}\frac{\delta(|y_{\perp}-v_{\perp}|\!-\!\tau)}{2\pi\tau}\frac{\delta(|y_{\perp}-\bar{v}_{\perp}|\!-\!\tau)}{2\pi\tau}\nonumber\\
&\times(1+\cos(\theta_{x-u}-\theta_{x-\bar{u}}))(1+\cos(\theta_{y-v}-\theta_{y-\bar{v}})),\label{tauEd2p}
\end{align}
\vspace{0.5cm}
\begin{align}
\langle\dot{\nu}(\tau,x_{\perp})\dot{\nu}(\tau,&\,y_{\perp})\rangle\!=\!\frac{g^4}{4}f^{abn}f^{cdn}f^{a'\!b'\!m}f^{c'\!d'\!m}\!\!\int\limits_{p,k}\int\limits_{\bar{p},\bar{k}}\int\limits_{u,v}\int\limits_{\bar{u},\bar{v}}\langle\alpha^{i,a}_u\alpha^{k,c}_{\bar{u}}\alpha^{i'\!\!,a'\!\!}_v\alpha^{k'\!\!,c'\!\!}_{\bar{v}}\rangle\langle\alpha^{j,b}_u\alpha^{l,d}_{\bar{u}}\alpha^{j'\!\!,b'\!\!}_v\alpha^{l'\!\!,d'\!\!}_{\bar{v}}\rangle\nonumber\\
&\times\!\!\left(\delta^{ij}\epsilon^{kl}J_0(p\tau)J_0(\bar{p}\tau)\!-\epsilon^{ij}\delta^{kl}\frac{p_{\perp}\!\cdot\!\bar{p}_{\perp}}{p\,\bar{p}}J_1(p\tau)J_1(\bar{p}\tau)\right)\!\!\left(\delta^{i'\!j'\!}\epsilon^{k'\!l'\!}J_0(k\tau)J_0(\bar{k}\tau)\!-\epsilon^{i'\!j'\!}\delta^{k'\!l'\!}\frac{k_{\perp}\!\cdot\!\bar{k}_{\perp}}{k\,\bar{k}}J_1(k\tau)J_1(\bar{k}\tau)\right)\nonumber\\
&\times\!e^{ip_{\perp}\!(x-u)_{\perp}}e^{ik_{\perp}\!(y-v)_{\perp}}e^{i\bar{p}_{\perp}\!(x-\bar{u})_{\perp}}e^{i\bar{k}_{\perp}\!(y-\bar{v})_{\perp}}\nonumber\\
=&\,\frac{g^4}{4}f^{abn}f^{cdn}f^{a'\!b'\!m}f^{c'\!d'\!m}\!\!\int\limits_{u,v}\int\limits_{\bar{u},\bar{v}}\langle\alpha^{i,a}_u\alpha^{k,c}_{\bar{u}}\alpha^{i'\!\!,a'\!\!}_v\alpha^{k'\!\!,c'\!\!}_{\bar{v}}\rangle\langle\alpha^{j,b}_u\alpha^{l,d}_{\bar{u}}\alpha^{j'\!\!,b'\!\!}_v\alpha^{l'\!\!,d'\!\!}_{\bar{v}}\rangle\frac{\delta(|x_{\perp}-u_{\perp}|\!-\!\tau)}{2\pi\tau}\frac{\delta(|x_{\perp}-\bar{u}_{\perp}|\!-\!\tau)}{2\pi\tau}\nonumber\\
&\times\frac{\delta(|y_{\perp}-v_{\perp}|\!-\!\tau)}{2\pi\tau}\frac{\delta(|y_{\perp}-\bar{v}_{\perp}|\!-\!\tau)}{2\pi\tau}(\delta^{ij}\epsilon^{kl}+\epsilon^{ij}\delta^{kl}\cos(\theta_{x-u}-\theta_{x-\bar{u}}))(\delta^{i'\!j'\!}\epsilon^{k'\!l'\!}+\epsilon^{i'\!j'\!}\delta^{k'\!l'\!}\cos(\theta_{y-v}-\theta_{y-\bar{v}})).
\label{taunud2p}
\end{align}
\end{widetext}
The physical interpretation of the results~\nr{tauEd2p} and~\nr{taunud2p} is quite clear in light of our discussion of the one-point function in \secref{sec:onept}. It consists of energy densities at points $x_\perp$, $y_\perp$ resulting from spherically expanding waves starting from points $u_\perp,\bar{u}_\perp,v_\perp,\bar{v}_\perp$. 

\begin{figure}
\centering
\includegraphics[width=0.48\textwidth]{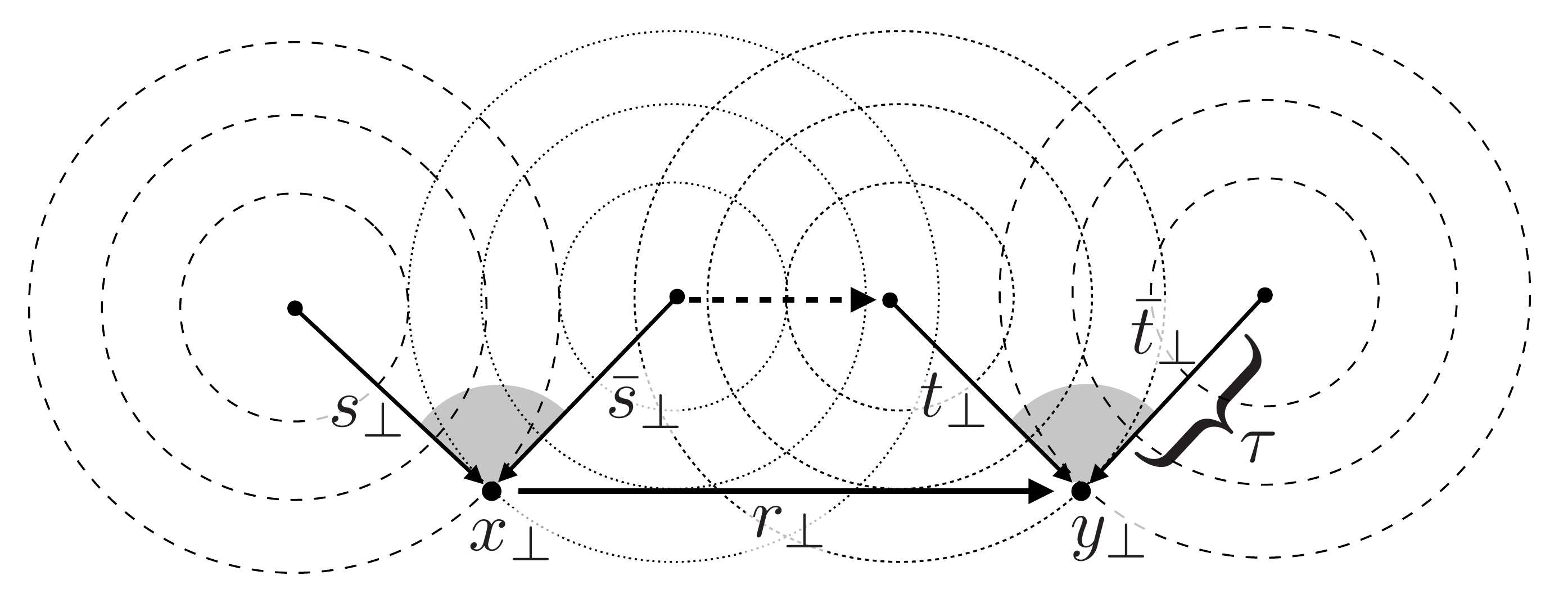}
\caption{Representation of an infinitesimal contribution to the two-point functions. For example here, the expanding spherical waves with dotted lines correspond to a two-point function $G_{1}((\bar{s}-t)_{\tau}-r_{\perp})$, where the distance argument of $G$ is denoted by the dashed arrow. The contribution from the other two-point function in the same nucleus  (spherical waves with dashed lines) is $G_{1}((s-\bar{t})_{\tau}-r_{\perp})$. The shaded angles control the interference effect.}
\label{waves2}
\end{figure}

As discussed in \secref{inco}, the building block of these correlators is the four-point function of the gluon fields, $\langle\alpha^{i,a}_u\alpha^{k,c}_{\bar{u}}\alpha^{i'\!\!,a'\!\!}_v\alpha^{k'\!\!,c'\!\!}_{\bar{v}}\rangle$. In the present work, we apply the glasma graph approximation in order to circumvent the complexity that this object presents.
Moreover, in the formulas presented below, we also adopt the GBW model, which significantly simplifies the resulting expressions.
After performing the variable changes $s_{\perp}\!=\!x_{\perp}-u_{\perp}$, $\bar{s}_{\perp}\!=x_{\perp}-\bar{u}_{\perp}$, $t_{\perp}\!=\!y_{\perp}-v_{\perp}$, $\bar{t}_{\perp}\!=y_{\perp}-\bar{v}_{\perp}$ and computing the analytically doable integrals, we obtain
\begin{widetext}
\begin{align}
\langle\varepsilon(\tau,x_{\perp})&\,\varepsilon(\tau,y_{\perp})\rangle\!=\!\frac{g^4}{8}N^2_c(N^2_c-1)\!\!\int_0^{2\pi}\frac{d\theta_s}{2\pi}\frac{d\theta_{\bar{s}}}{2\pi}\frac{d\theta_t}{2\pi}\frac{d\theta_{\bar{t}}}{2\pi}(1+\cos(\theta_s-\theta_{\bar{s}}))(1+\cos(\theta_t-\theta_{\bar{t}}))\nonumber\\
\times\!\bigg[&\bigg((N_c^2-1)G_1((s-\bar{s})_{\tau})G_1((t-\bar{t})_{\tau})G_2((s-\bar{s})_{\tau})G_2((t-\bar{t})_{\tau})\nonumber\\
&+2G_1((s-\bar{s})_{\tau})G_1((t-\bar{t})_{\tau})G_2((s-t)_{\tau}-r_{\perp})G_2((\bar{s}-\bar{t})_{\tau}-r_{\perp})\nonumber\\
&+G_1((s-t)_{\tau}-r_{\perp})G_1((\bar{s}-\bar{t})_{\tau}-r_{\perp})G_2((s-t)_{\tau}-r_{\perp})G_2((\bar{s}-\bar{t})_{\tau}-r_{\perp})\bigg)\!+\!\bigg(\!1\leftrightarrow2\!\bigg)\bigg],\label{tauEd2pp}
\end{align}
\begin{align}
\langle&\dot{\nu}(\tau,x_{\perp})\dot{\nu}(\tau,y_{\perp})\rangle\!=\!\frac{g^4}{64}N^2_c(N^2_c-1)\!\!\int_0^{2\pi}\frac{d\theta_s}{2\pi}\frac{d\theta_{\bar{s}}}{2\pi}\frac{d\theta_t}{2\pi}\frac{d\theta_{\bar{t}}}{2\pi}(1+\cos(\theta_s-\theta_{\bar{s}}))(1+\cos(\theta_t-\theta_{\bar{t}}))\nonumber\\
\bigg[&\bigg(G_1((s-t)_{\tau}-r_{\perp})G_1((\bar{s}-\bar{t})_{\tau}-r_{\perp})G_2((s-\bar{t})_{\tau}-r_{\perp})G_2((\bar{s}-t)_{\tau}-r_{\perp})\nonumber\\
&+2G_1((s-t)_{\tau}-r_{\perp})G_1((\bar{s}-\bar{t})_{\tau}-r_{\perp})G_2((s-t)_{\tau}-r_{\perp})G_2((\bar{s}-\bar{t})_{\tau}-r_{\perp})
\!\bigg)\!+\!\bigg(\!1\leftrightarrow2\!\bigg)\bigg],\label{taun2pp}
\end{align}
\end{widetext}
where $G_{1,2}$ are the unpolarized gluon distributions defined by \eqref{Fields2pf}. The general glasma graph expressions of Eqs.\ (\ref{tauEd2pp}) and (\ref{taun2pp}) including the linearly polarized distribution are shown in Appendix~\ref{2pedgg}.
Here we have defined $(s\!-\!t)_{\tau}\!\equiv\!\tau(\hat{s}_{\perp}\!-\!\hat{t}_{\perp})$. This is a subtraction of two vectors $s$ and $t$ whose moduli have been replaced by $\tau$ as a result of the integration of the delta functions. Note that this does not represent a transverse shift of length $\tau$; instead, the value of $|(s-t)_{\tau}|$ depends on the integration variables $\theta_s$, $\theta_t$. 
Fig.\!~\ref{waves2} provides a graphical interpretation of the transverse coordinate structure corresponding to this expression.

The glasma graph approximation allows us to distinguish between contributions stemming from ``connected'' or ``disconnected'' correlators, defined as
\begin{align}
\langle &\,\alpha^{i,a}_x\alpha^{j,b}_x\alpha^{k,c}_y\alpha^{l,d}_y\rangle=\overbrace{\langle \alpha^{i,a}_x\alpha^{j,b}_x\rangle\langle \alpha^{k,c}_y\alpha^{l,d}_y\rangle}^{\scaleto{\text{disconnected}}{4pt}}\nonumber\\
&+\underbrace{\langle  \alpha^{i,a}_x\alpha^{k,c}_y\rangle\langle \alpha^{j,b}_x\alpha^{l,d}_y\rangle+\langle \alpha^{i,a}_x\alpha^{l,d}_y\rangle\langle\alpha^{j,b}_x\alpha^{k,c}_y\rangle}_{\scaleto{\text{connected}}{4pt}}.
\end{align}
This is the usual terminology employed in previous calculations of correlation functions at $\tau\!=\!0^+$. Although we have not explicitly made this distinction here, in Eqs.\ (\ref{tauEd2pp}) and (\ref{taun2pp}) one can identify the connected contributions as those that depend on the correlation distance $r_{\perp}$. Another remarkable aspect about the previous formulas lies in the overall interference factor, which results simply from squaring that of the one-point function $\langle\varepsilon\rangle$. This feature is maintained in the general expressions obtained under the glasma graph approximation (see Appendix~\ref{2pedgg}) and, at least in the case of the energy density, also in the full nonlinear calculation. 

\begin{figure*}
\centering
\subfigure{\label{rgbw}\includegraphics[width=0.49\textwidth]{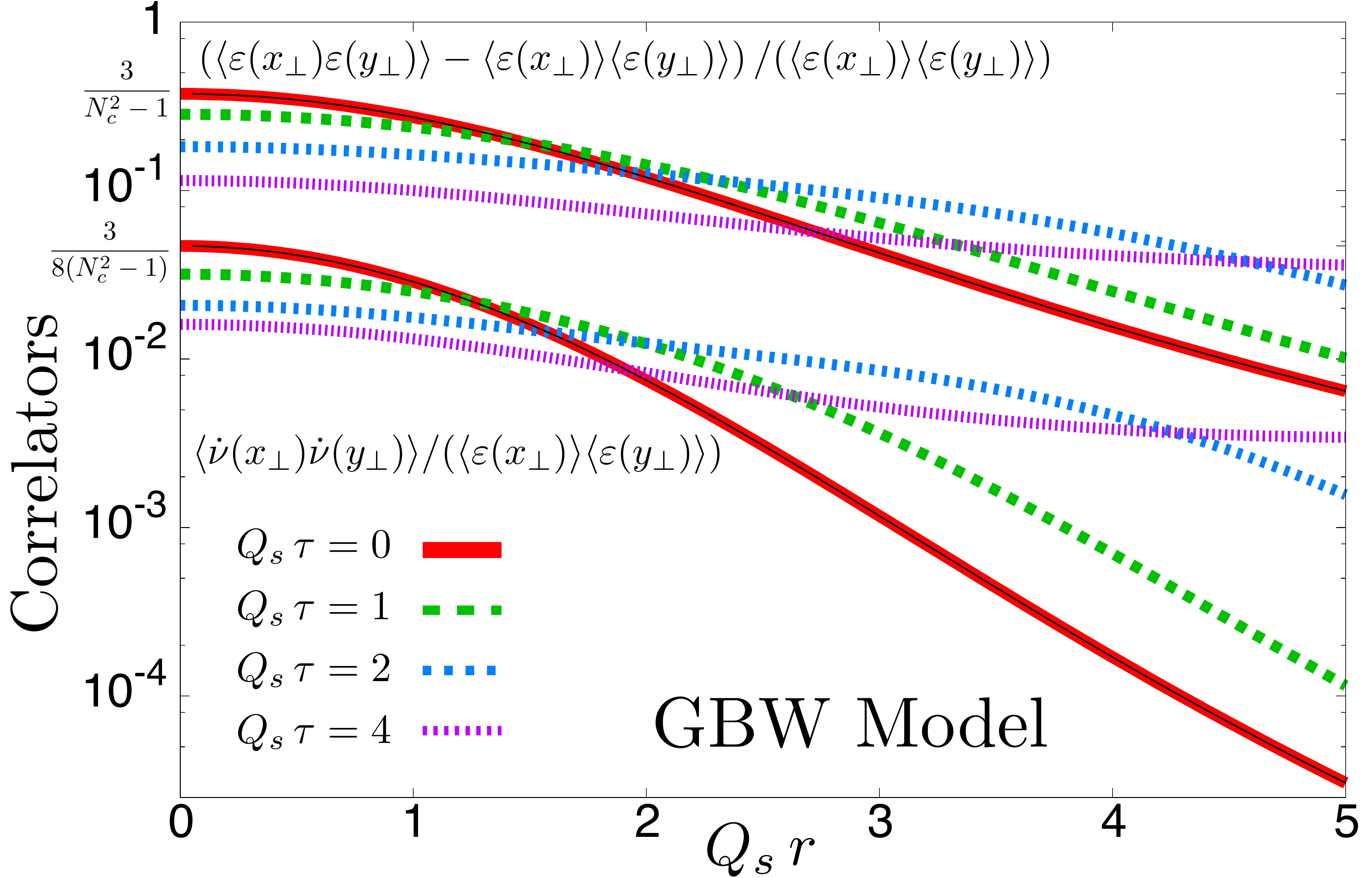}}
\subfigure{\label{rmv}\includegraphics[width=0.49\textwidth]{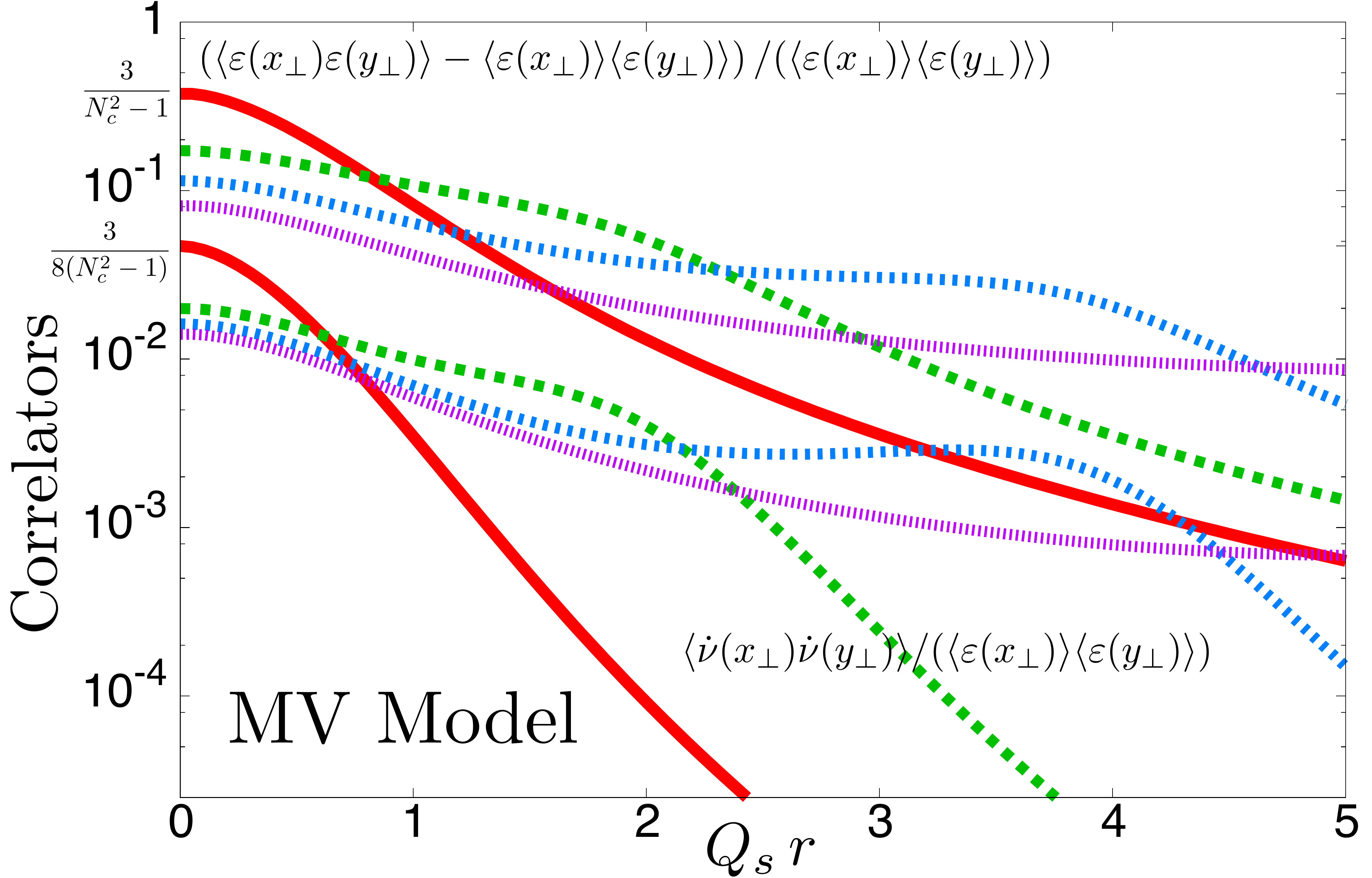}}
\caption{
Correlation functions of the energy density and the divergence of the Chern-Simons current at different values of $\tau$ for the GBW model (left plot) and the MV model (right plot). The thin black lines in the left plot correspond to Eqs.\ (\ref{2p0e}) and (\ref{2p0n}).}
\label{Result3}
\end{figure*}

As we have explicitly integrated out the highly oscillatory Bessel functions that originally contained the $\tau$ dependence, the resulting expressions turn out to be relatively simple to solve with numerical methods. We did this for different multiples of $\tau\!=\!1/Q_s$ applying the global adaptive method built in \textit{Mathematica}.
The results are shown in \figref{Result3}. We can see how linear Yang-Mills evolution has a qualitatively similar effect on the glasma properties in both models: increasing the magnitude of long-distance fluctuations while decreasing that of the short-distance ones.
Although the definition of a correlation length is somewhat shaky in this context (as these correlations decay following a power-law tail rather than exponentially), this trend could be roughly described as a correlation length growth, which is one of the effects expected to take place in any system that is approaching the hydrodynamical regime. 

Written in the form \nr{tauEd2pp} and \nr{taun2pp} [or \nr{tauEd2ppGEN}, \nr{taun2ppGEN}], it is not manifest that our expressions indeed reproduce the $\tau\!=\!0^+$ case obtained in previous works. This can be checked analytically by
 noting that setting $\tau\!=\!0$ in Eqs.\ \nr{tauEd2ppGEN} and \nr{taun2ppGEN} enables us to integrate out the interference factors straightforwardly. By doing so, one directly obtains the formulas presented in Sec.\ 3.3 of \cite{Lappi:2017skr}. We show the result from numerically evaluating  the $\tau\!=\!0^+$ results  \nr{2p0e} and \nr{2p0n} in  \figref{rgbw}.
The corresponding figures for the MV model are shown in \figref{rmv}. Note that, as we adopted a different running coupling prescription, the curves corresponding to the MV model at $\tau\!=\!0^+$ [thick lines in \figref{rmv}] do not exactly match the ones presented in \cite{Lappi:2017skr}. As a check, we verified that our results do agree when we include running coupling corrections in the same way. Also, we repeated our calculations using the MV model with fixed coupling and compared them to the curves shown in \figref{rmv}. The comparison between these results (included in Appendix~\ref{norc}) illustrates how our running coupling prescription successfully preserves nontrivial aspects of the MV model both in coordinate and momentum space.

Fig.\!~\ref{Result3} shows that the decay of the correlators, both in space and time, is significantly more pronounced under the MV model. This is evident from the steepness of the curves and from their decreasing normalization as we move to larger values of $\tau$. Indeed, in the curves corresponding to the MV model, we observe a relatively large drop in the values at $Q_{s}r\!=\!0$ when we evolve from $Q_{s}\tau\!=\!0$ to $Q_{s}\tau\!=\!1$. This trend is, however, stabilized for later values of $\tau$, decreasing at a similar pace as the GBW model result. This is somewhat similar to the steeper decay of the correlator with $Q_{s}r$ in the MV model. We attribute this effect to the way in which $r$ and $\tau$ enter our formulas; both as arguments of the gluon distributions $G_{1,2}$, $h_{1,2}$, as well as other functions defined in Appendix~\ref{2pedgg}. It is thus reasonable that our curves exhibit a similar (although not identical) behavior with $Q_{s}r$ and $Q_{s}\tau$.

\section{Conclusions}

In this work, we have presented an analytical calculation of one- and two-point correlators of energy density and axial charge at finite proper times. These objects characterize the average and fluctuations of energy density and axial charge deposited throughout the initial stage of HICs, during which a classical description of the system is appropriate. In our calculation, we assume a free-field evolution of the gluon fields. This setup has been previously applied in the literature for the description of dilute-dense collisions \cite{Dumitru:2001ux,Kovchegov:2005ss,Kovchegov:2005kn,McLerran:2016snu,Lappi:2017skr}. In these works, the initial conditions are expanded in powers of the weak field describing the gluon content of the dilute nucleus. To the lowest order in this expansion, the dynamics of the system are described by the linearized Yang-Mills equations. In the present paper, we propose that the same equations can be applied to the evolution of the full initial conditions, which encode the saturation features of the system. Our claim is supported by the dominance of the high-momentum modes of a power series in proper time, observed in \cite{Fries:2006pv,Fries:2007iy,Fujii:2008km,Chen:2015wia}. The UV divergences carried by all terms of this series are effectively resummed by considering the higher order derivatives of the Yang-Mills equations \cite{Fujii:2008km}. We argue that nonlinear corrections to this approach are negligible when focusing on quantities dominated by large momenta, such as the energy density and the divergence of the Chern-Simons current.

In this paper, we evaluate the initial conditions in the glasma graph approximation, which assumes a Gaussian-like decomposition of four-point correlators into two-point correlators for the gluon fields. In order to obtain quantitative results, we adopt the GBW and MV models, including running coupling corrections in the last case. Our running coupling regularization of the MV model yields a finite result in the UV limit, while preserving a perturbative power-law tail at large transverse momenta of the unintegrated gluon distribution (a signature feature of the MV model). This running coupling  approach differs from that of previous works \cite{Lappi:2017skr}, hence yielding slightly different curves at $\tau\!=\!0^+$. In the GBW model, in contrast, the unintegrated gluon distribution falls off as a Gaussian at large transverse momenta.  Nevertheless, for the energy density correlator, the two parametrizations give qualitatively similar results. 

Yang-Mills evolution was analytically computed for all correlators up until a proper time $\tau\!=\!4/Q_s$.
The evolution of the average energy density deposited on a given point can be described through a dimensionless dilution factor, which accounts for both the attenuation and the interference of the expanding plane waves that describe the gluon propagation in the plane transverse to the collision axis. We observe that in the MV model this dilution effect is noticeably more rapid than for the GBW dipole correlator. This can be naturally understood as the effect of a larger relative contribution of fast-evolving large transverse momentum modes in the MV model, where they are only suppressed  as a power law and not a Gaussian. The effect of time evolution over the considered two-point correlators can be described as an elongation of the correlation length, a trend that is suggestive of a transition toward the hydrodynamical regime.

The glasma graph approximation has been adopted here merely for calculational convenience, and we expect it to be relatively well satisfied apart from the longest coordinate separations $r\gtrsim 1/Q_s$. However, our approach for computing the time dependence is much more general, and can be applied to any initial color fields, as long as the four-point function of the Weizs\"acker-Williams fields of the colliding nuclei can be computed. 
A reliable application of our results into the study of eccentricity harmonics will require doing this and reevaluating our $\tau$-dependent correlators in the full MV model, i.e.\ beyond the glasma graph approximation. This is due to the sensitivity of the mean-squared eccentricity fluctuations to large-scale correlations ($Q_sr\!>\!1$), a regime that lies outside the validity region of the glasma graph approximation.

\section{acknowledgements}
The authors thank S\"oren Schlichting for providing useful insight on the running coupling prescription presented in \cite{Lappi:2017skr} and discussions on the gauge dependence. P.G.-R.\ thanks Cyrille Marquet and Heikki M\"antysaari for their useful feedback during the development of this work. T.L.\ thanks Matt Sievert for pertinent questions on the time dependence of the glasma energy density.
This work was supported by the Academy of Finland, Project No.\ 321840 and under the European Union's Horizon 2020 research and innovation programme by the European Research Council (ERC, Grant No.\ ERC-2015-CoG-681707) and by the H2020 Research Infrastructures STRONG-2020 project (Grant No.\ 824093). The content of this paper does not reflect the official opinion of the European Union and responsibility for the information and views expressed therein lies entirely with the authors. 

\appendix
\section{The divergence of the Chern-Simons current}\label{CME}
The chiral anomaly of QCD induces a transformation of left- into right-handed quarks, or equivalently, a generation of axial charge $N_5$,
\begin{align}
\frac{dN_5}{dt}=\,&\frac{d(N_R-N_L)}{dt}\nonumber\\
=\,&-\frac{g^2N_f}{8\pi^2}\int d^3x\,\text{Tr}\left\{F_{\mu\nu}(x)\tilde{F}^{\mu\nu}(x)\right\},\label{DCSf}
\end{align}
where $\tilde{F}^{\mu\nu}\!=\!\frac{1}{2}\epsilon^{\mu\nu\rho\sigma}F_{\rho\sigma}$ is the dual of the field strength tensor and $N_f$ represents the number of flavors. In this formula we can see how the production rate of axial charge is directly related to the properties of the gauge fields entering the right-hand side of the equation. This term implicitly includes various contributions to axial charge production, one of them stemming directly from the topological structure of QCD.

We can define distinct topological classes of gauge field configurations labeled by the following quantity:
\begin{align}
Q_{\text{w}}=\frac{g^2}{16\pi^2}\int\!d^4x\,\text{Tr}\left\{F_{\mu\nu}(x)\tilde{F}^{\mu\nu}(x)\right\},
\end{align}
a topological invariant known as winding number. These classes are separated by potential barriers with heights of order $\Lambda_{\text{QCD}}$, which suppress the $Q_{\text{w}}$ fluctuations at low temperature. However, that is not the case in a high temperature medium such as the QGP. At $T\!\sim\!\Lambda_{\text{QCD}}$, $Q_{\text{w}}$ fluctuations happen with a rate that can be directly related to $dN_5/dt$. This relation is expressed in terms of the Chern-Simons current, defined as
\begin{align}
K^{\mu}=\epsilon^{\mu\nu\rho\sigma}A^{a}_{\nu}\left(F^{a}_{\rho\sigma}+\frac{g}{3}A^{b}_{\rho}A^{c}_{\sigma}\right),
\end{align}
and whose divergence reads
\begin{align}
\partial_{\mu}K^{\mu}=-\frac{1}{4}\text{Tr}\left\{F_{\mu\nu}(x)\tilde{F}^{\mu\nu}(x)\right\}\equiv\dot{\nu}(x).
\end{align}
It is convenient to rewrite \eqref{DCSf} in terms of the divergence of the Chern-Simons current,
\begin{align}
\frac{dN_5}{dt}=\frac{g^2N_f}{2\pi^2}\int d^3x\,\dot{\nu}(x).
\end{align}
Based on this relation, in the present work, we take $\dot{\nu}$ as the fundamental object controlling axial charge generation. This effect may manifest experimentally in off-central HICs, where large background electromagnetic fields are generated \cite{Skokov:2009qp}. These fields, when in the presence of deconfined chirally unbalanced matter, induce a separation of charges that gives rise to an electric dipole along the direction of angular momentum. This is known as the chiral magnetic effect, and it translates into nontrivial azimuthal correlations in the hadron spectrum \cite{Kharzeev:2001ev,Kharzeev:2007jp}. This signal, however, is obscured by large background effects that produce similar back-to-back correlations. Reducing this uncertainty is currently a major goal of the high energy QCD community, both on the experimental \cite{Voloshin:2004vk,Bzdak:2012ia,Wen:2016zic,Xu:2017qfs,Huang:2018duz} and the theoretical side \cite{Schlichting:2010qia,Lappi:2017skr,Guerrero-Rodriguez:2019ids,Hammelmann:2019vwd}.
\begin{figure*}
\centering
\subfigure{\label{Result_norc1}\includegraphics[width=0.49\textwidth]{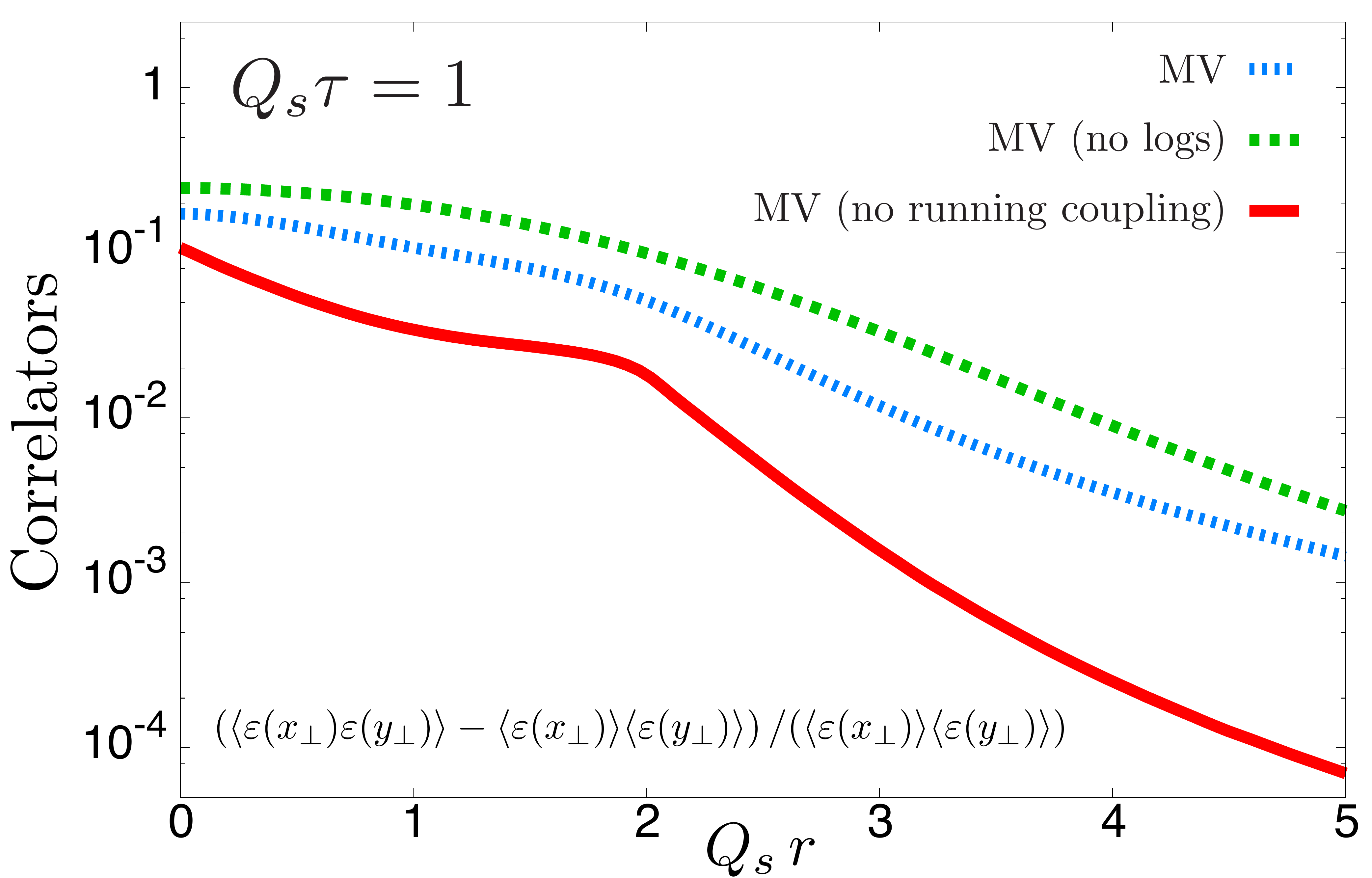}}
\subfigure{\label{Result_norc2}\includegraphics[width=0.49\textwidth]{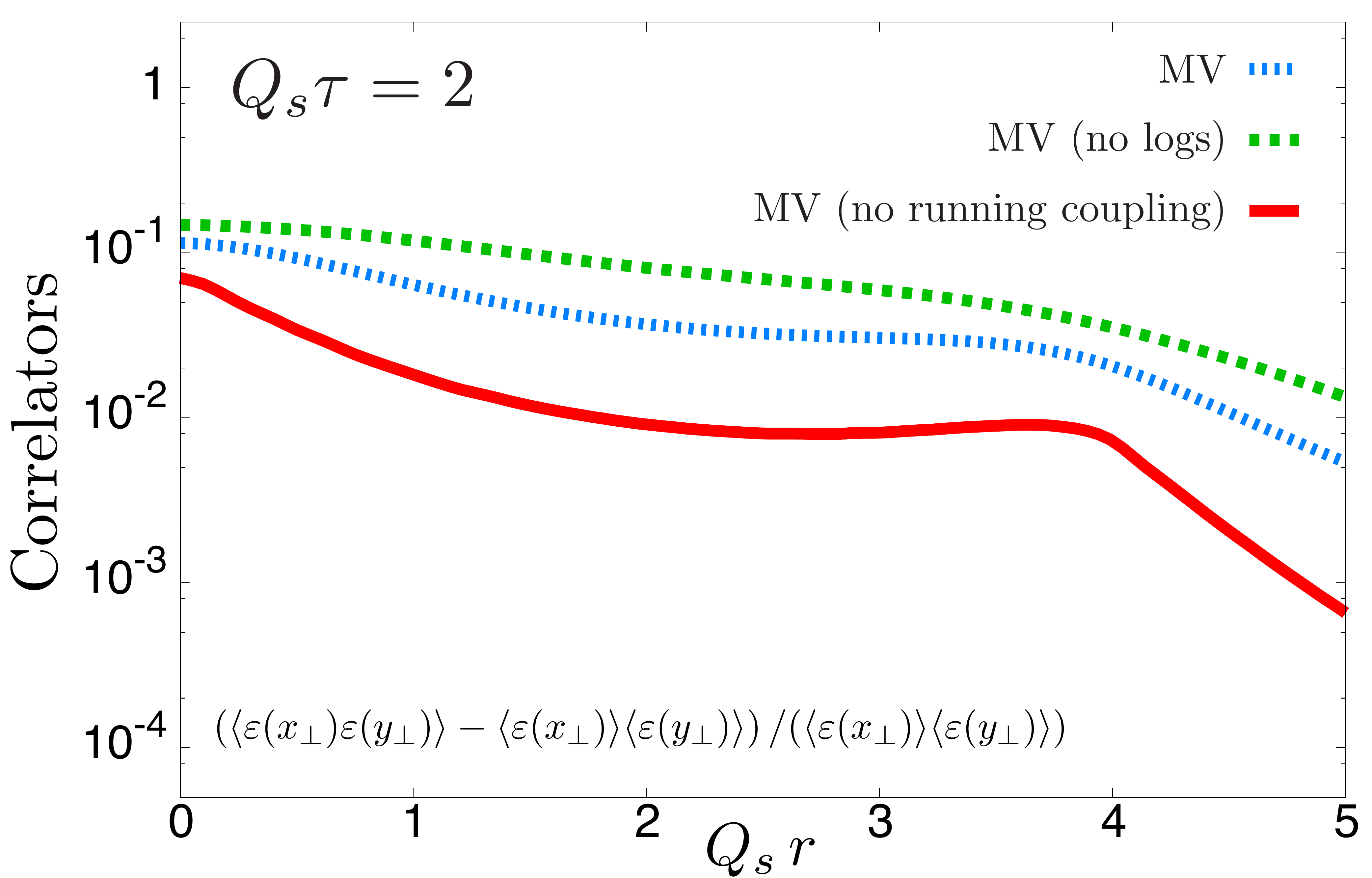}}
\caption{Correlation functions of the energy density computed under different running coupling prescriptions at different proper times.}
\label{Result_norc}
\end{figure*}
In this context, it is of paramount importance to constrain the dynamical origin of CME signatures. The calculation of $\dot{\nu}$ correlators presented in this paper accounts for the $\tau$ evolution of potential contributions emerging from event-by-event color charge fluctuations during the glasma phase. This source, although unrelated to the topological structure of QCD, can have a sizable effect on final state observables related to \textit{CP} violation. The exploration of this possibility is left for future phenomenological studies.

\begin{widetext}
\section{Two-point functions in the glasma graph approximation}\label{2pedgg}
In the following appendix, we display the general expressions of the correlators of the energy density and axial charge deposited in two points $x_{\perp}$ and $y_{\perp}$ of the transverse plane at a proper time $\tau$ within the glasma graph approximation as follows:
\begin{align}
\langle&\,\varepsilon(\tau,x_{\perp})\varepsilon(\tau,y_{\perp})\rangle\!=\!\frac{g^4}{8}N^2_c(N^2_c-1)\!\!\int_0^{2\pi}\frac{d\theta_s}{2\pi}\frac{d\theta_{\bar{s}}}{2\pi}\frac{d\theta_t}{2\pi}\frac{d\theta_{\bar{t}}}{2\pi}(1+\cos(\theta_s-\theta_{\bar{s}}))(1+\cos(\theta_t-\theta_{\bar{t}}))\nonumber\\
&\times\!\bigg[\bigg((N_c^2-1)G_1((s-\bar{s})_{\tau})G_1((t-\bar{t})_{\tau})G_2((s-\bar{s})_{\tau})G_2((t-\bar{t})_{\tau})\nonumber\\
&+2G_1((s-\bar{s})_{\tau})G_1((t-\bar{t})_{\tau})G_2((s-t)_{\tau}-r_{\perp})G_2((\bar{s}-\bar{t})_{\tau}-r_{\perp})\nonumber\\
&+G_1((s-t)_{\tau}-r_{\perp})G_1((\bar{s}-\bar{t})_{\tau}-r_{\perp})G_2((s-t)_{\tau}-r_{\perp})G_2((\bar{s}-\bar{t})_{\tau}-r_{\perp})\nonumber\\
&+2G_1((s-\bar{s})_{\tau})G_1((t-\bar{t})_{\tau})h_2((s-t)_{\tau}-r_{\perp})h_2((\bar{s}-\bar{t})_{\tau}-r_{\perp})f\Big((s-t)_{\tau}-r_{\perp},(\bar{s}-\bar{t})_{\tau}-r_{\perp}\Big)\nonumber\\
&+G_1((s-t)_{\tau}-r_{\perp})G_1((\bar{s}-\bar{t})_{\tau}-r_{\perp})h_2((s-\bar{t})_{\tau}-r_{\perp})h_2((\bar{s}-t)_{\tau}-r_{\perp})f\Big((s-\bar{t})_{\tau}-r_{\perp},(\bar{s}-t)_{\tau}-r_{\perp}\Big)\nonumber\\
&+h_1((s-t)_{\tau}-r_{\perp})h_1((\bar{s}-\bar{t})_{\tau}-r_{\perp})h_2((s-t)_{\tau}-r_{\perp})h_2((\bar{s}-\bar{t})_{\tau}-r_{\perp})\bigg)\!+\!\bigg(\!1\leftrightarrow2\!\bigg)\bigg],\label{tauEd2ppGEN}
\end{align}
\begin{align}
\langle&\,\dot{\nu}(\tau,x_{\perp})\dot{\nu}(\tau,y_{\perp})\rangle\!=\!\frac{g^4}{64}N^2_c(N^2_c-1)\!\!\int_0^{2\pi}\frac{d\theta_s}{2\pi}\frac{d\theta_{\bar{s}}}{2\pi}\frac{d\theta_t}{2\pi}\frac{d\theta_{\bar{t}}}{2\pi}(1+\cos(\theta_s-\theta_{\bar{s}}))(1+\cos(\theta_t-\theta_{\bar{t}}))\nonumber\\
&\times\!\bigg[\bigg(G_1((s-t)_{\tau}-r_{\perp})G_1((\bar{s}-\bar{t})_{\tau}-r_{\perp})G_2((s-\bar{t})_{\tau}-r_{\perp})G_2((\bar{s}-t)_{\tau}-r_{\perp})\nonumber\\
&+2G_1((s-t)_{\tau}-r_{\perp})G_1((\bar{s}-\bar{t})_{\tau}-r_{\perp})G_2((s-t)_{\tau}-r_{\perp})G_2((\bar{s}-\bar{t})_{\tau}-r_{\perp})\nonumber\\
&-2h_1((s-t)_{\tau}-r_{\perp})h_1((\bar{s}-\bar{t})_{\tau}-r_{\perp})h_2((s-t)_{\tau}-r_{\perp})h_2((\bar{s}-\bar{t})_{\tau}-r_{\perp})\nonumber\\
&+4G_1((s-t)_{\tau}-r_{\perp})G_1((\bar{s}-\bar{t})_{\tau}-r_{\perp})h_2((s-\bar{s})_{\tau})h_2((t-\bar{t})_{\tau})f\Big((s-\bar{s})_{\tau},(t-\bar{t})_{\tau}\Big)\nonumber\\
&-h_1((s-t)_{\tau}-r_{\perp})h_1((\bar{s}-\bar{t})_{\tau}-r_{\perp})h_2((s-\bar{t})_{\tau}-r_{\perp})h_2((\bar{s}-t)_{\tau}-r_{\perp})g_1(\tau,r,\theta)\nonumber\\
&-4h_1((s-t)_{\tau}-r_{\perp})h_1((\bar{s}-\bar{t})_{\tau}-r_{\perp})h_2((s-\bar{s})_{\tau})h_2((t-\bar{t})_{\tau})g_2(\tau,r,\theta)\bigg)\!+\!\bigg(\!1\leftrightarrow2\!\bigg)\bigg]\label{taun2ppGEN}.
\end{align}
For the sake of compactness, we have defined the following functions:
\begin{align}
f(x_{\perp},y_{\perp})=&\cos\left(2(\theta_x-\theta_y)\right)=2\left(\frac{x_{\perp}\cdot y_{\perp}}{xy}\right)^2-1,\\
g_1(\tau,r,\theta)=&f((s-t)_{\tau}-r_{\perp},(s-\bar{t})_{\tau}-r_{\perp})f((\bar{s}-\bar{t})_{\tau}-r_{\perp},(\bar{s}-t)_{\tau}-r_{\perp})\nonumber\\
&+f((s-t)_{\tau}-r_{\perp},(\bar{s}-t)_{\tau}-r_{\perp})f((\bar{s}-\bar{t})_{\tau}-r_{\perp},(s-\bar{t})_{\tau}-r_{\perp})\nonumber\\
&-f((s-t)_{\tau}-r_{\perp},(\bar{s}-\bar{t})_{\tau}-r_{\perp})f((s-\bar{t})_{\tau}-r_{\perp},(\bar{s}-t)_{\tau}-r_{\perp}),\\
g_2(\tau,r,\theta)=&f((s-\bar{s})_{\tau},(s-t)_{\tau}-r_{\perp})f((t-\bar{t})_{\tau},(\bar{s}-\bar{t})_{\tau}-r_{\perp})\nonumber\\
&+f((s-\bar{s})_{\tau},(\bar{s}-\bar{t})_{\tau}-r_{\perp})f((t-\bar{t})_{\tau},(s-t)_{\tau}-r_{\perp})\nonumber\\
&-f((s-\bar{s})_{\tau},(t-\bar{t})_{\tau})f((s-t)_{\tau}-r_{\perp},(\bar{s}-\bar{t})_{\tau}-r_{\perp}).
\end{align}
The indicated integrals were performed numerically with \textit{Mathematica} to obtain the curves shown in Fig.\!~\ref{Result3}.
\end{widetext}

\section{Two-point function of energy density in the MV model with fixed coupling}\label{norc}
The energy density of the glasma fields in the MV model is finite for all $\tau\!>\!0$, which allows to compute the $\tau$ evolution of its correlators without resorting to any UV regularization scheme. Under this approach, we could compare results in the GBW and fixed coupling MV models exclusively at finite $\tau$ values, or alternatively, regularize the UV divergence only at $\tau\!=\!0^+$. Instead of that, in this work, we adopt the running coupling prescription described in \secref{incoMV}, which defines a variant of the MV model that we consistently apply throughout the whole $\tau$ range considered. This ansatz also gives rise to less numerically demanding integrals.

However, the fixed coupling MV model yields some nontrivial results that are worth remarking. In \figref{Result_norc}, one can see that the correlator corresponding to the MV model with no running coupling (thick curve) features a significantly steeper falloff at large transverse separations.
The correlator enters this fast decay region somewhat abruptly (at $r \approx 2/Q_s$ for $\tau\!=\!1/Q_s$ and at $r \approx 4/Q_s$ for $\tau\!=\!2/Q_s$), giving rise to a nontrivial profile that hints at the formation of a relatively flat plateau at short correlation distances. Remarkably, this characteristic shape is also visible in the correlators obtained under our running coupling prescription, although it appears less prominently and at later proper times (around $\tau\!=\!2/Q_s$, as can be seen in \figref{Result_norc2}).

In \figref{Result_norc}, we also display the correlator obtained by applying the running coupling prescription \eqref{rc1} to all the coupling constants of Eqs.~\nr{eq:mvG} and~\nr{eq:mvH} (green dashed curve). This curve, labeled as MV (no logs), shows none of the features mentioned above, being qualitatively identical to the GBW result. The comparison between these correlators supports our particular choice for the running coupling prescription, as it was made with the aim of preserving (to the extent possible) the nontrivial aspects of the fixed coupling MV model while simultaneously providing finite results at $\tau\!=\!0^+$.

\bibliographystyle{JHEP-2modlong.bst}

\begin{thebibliography}{10}

\bibitem{Luzum:2013yya}
M.~Luzum and H.~Petersen, {\it {Initial State Fluctuations and Final State
  Correlations in Relativistic Heavy-Ion Collisions}},
  \href{http://dx.doi.org/10.1088/0954-3899/41/6/063102}{{\em J. Phys. G} {\bf
  41} (2014) 063102} [\href{http://arXiv.org/abs/1312.5503}{{\tt
  arXiv:1312.5503 [nucl-th]}}].

\bibitem{Gelis:2010nm}
F.~Gelis, E.~Iancu, J.~Jalilian-Marian and R.~Venugopalan, {\it {The Color
  Glass Condensate}},
  \href{http://dx.doi.org/10.1146/annurev.nucl.010909.083629}{{\em Ann. Rev.
  Nucl. Part. Sci.} {\bf 60} (2010) 463}
  [\href{http://arXiv.org/abs/1002.0333}{{\tt arXiv:1002.0333 [hep-ph]}}].

\bibitem{Lappi:2010ek}
T.~Lappi, {\it {Small x physics and RHIC data}},
  \href{http://dx.doi.org/10.1142/S0218301311017302}{{\em Int. J. Mod. Phys. E}
  {\bf 20} (2011) 1} [\href{http://arXiv.org/abs/1003.1852}{{\tt
  arXiv:1003.1852 [hep-ph]}}].

\bibitem{Albacete:2014fwa}
J.~L. Albacete and C.~Marquet, {\it {Gluon saturation and initial conditions
  for relativistic heavy ion collisions}},
  \href{http://dx.doi.org/10.1016/j.ppnp.2014.01.004}{{\em Prog. Part. Nucl.
  Phys.} {\bf 76} (2014) 1} [\href{http://arXiv.org/abs/1401.4866}{{\tt
  arXiv:1401.4866 [hep-ph]}}].

\bibitem{McLerran:1993ni}
L.~D. McLerran and R.~Venugopalan, {\it {Computing quark and gluon distribution
  functions for very large nuclei}},
  \href{http://dx.doi.org/10.1103/PhysRevD.49.2233}{{\em Phys. Rev. D} {\bf 49}
  (1994) 2233} [\href{http://arXiv.org/abs/hep-ph/9309289}{{\tt
  arXiv:hep-ph/9309289}}].

\bibitem{McLerran:1993ka}
L.~D. McLerran and R.~Venugopalan, {\it {Gluon distribution functions for very
  large nuclei at small transverse momentum}},
  \href{http://dx.doi.org/10.1103/PhysRevD.49.3352}{{\em Phys. Rev. D} {\bf 49}
  (1994) 3352} [\href{http://arXiv.org/abs/hep-ph/9311205}{{\tt
  arXiv:hep-ph/9311205}}].

\bibitem{McLerran:1994vd}
L.~D. McLerran and R.~Venugopalan, {\it {Green's functions in the color field
  of a large nucleus}},  \href{http://dx.doi.org/10.1103/PhysRevD.50.2225}{{\em
  Phys. Rev. D} {\bf 50} (1994) 2225}
  [\href{http://arXiv.org/abs/hep-ph/9402335}{{\tt arXiv:hep-ph/9402335}}].

\bibitem{Lappi:2006fp}
T.~Lappi and L.~McLerran, {\it {Some features of the glasma}},
  \href{http://dx.doi.org/10.1016/j.nuclphysa.2006.04.001}{{\em Nucl. Phys. A}
  {\bf 772} (2006) 200} [\href{http://arXiv.org/abs/hep-ph/0602189}{{\tt
  arXiv:hep-ph/0602189}}].

\bibitem{Dusling:2010rm}
K.~Dusling, T.~Epelbaum, F.~Gelis and R.~Venugopalan, {\it {Role of quantum
  fluctuations in a system with strong fields: Onset of hydrodynamical flow}},
  \href{http://dx.doi.org/10.1016/j.nuclphysa.2010.11.009}{{\em Nucl. Phys. A}
  {\bf 850} (2011) 69} [\href{http://arXiv.org/abs/1009.4363}{{\tt
  arXiv:1009.4363 [hep-ph]}}].

\bibitem{Gelis:2015gza}
F.~Gelis, {\it {Initial state and thermalization in the Color Glass Condensate
  framework}},  \href{http://dx.doi.org/10.1142/S0218301315300088}{{\em Int. J.
  Mod. Phys. E} {\bf 24} (2015) 1530008}
  [\href{http://arXiv.org/abs/1508.07974}{{\tt arXiv:1508.07974 [hep-ph]}}].

\bibitem{Berges:2020fwq}
J.~Berges, M.~P. Heller, A.~Mazeliauskas and R.~Venugopalan, {\it
  {Thermalization in QCD: theoretical approaches, phenomenological
  applications, and interdisciplinary connections}},
  \href{http://arXiv.org/abs/2005.12299}{{\tt arXiv:2005.12299 [hep-th]}}.

\bibitem{Kurkela:2016vts}
A.~Kurkela, {\it {Initial state of Heavy-Ion Collisions: Isotropization and
  thermalization}},
  \href{http://dx.doi.org/10.1016/j.nuclphysa.2016.01.069}{{\em Nucl. Phys. A}
  {\bf 956} (2016) 136} [\href{http://arXiv.org/abs/1601.03283}{{\tt
  arXiv:1601.03283 [hep-ph]}}].

\bibitem{Kurkela:2014tea}
A.~Kurkela and E.~Lu, {\it {Approach to Equilibrium in Weakly Coupled
  Non-Abelian Plasmas}},
  \href{http://dx.doi.org/10.1103/PhysRevLett.113.182301}{{\em Phys. Rev.
  Lett.} {\bf 113} (2014) 182301} [\href{http://arXiv.org/abs/1405.6318}{{\tt
  arXiv:1405.6318 [hep-ph]}}].

\bibitem{Kurkela:2015qoa}
A.~Kurkela and Y.~Zhu, {\it {Isotropization and hydrodynamization in weakly
  coupled heavy-ion collisions}},
  \href{http://dx.doi.org/10.1103/PhysRevLett.115.182301}{{\em Phys. Rev.
  Lett.} {\bf 115} (2015) 182301} [\href{http://arXiv.org/abs/1506.06647}{{\tt
  arXiv:1506.06647 [hep-ph]}}].

\bibitem{Keegan:2016cpi}
L.~Keegan, A.~Kurkela, A.~Mazeliauskas and D.~Teaney, {\it {Initial conditions
  for hydrodynamics from weakly coupled pre-equilibrium evolution}},
  \href{http://dx.doi.org/10.1007/JHEP08(2016)171}{{\em JHEP} {\bf 08} (2016)
  171} [\href{http://arXiv.org/abs/1605.04287}{{\tt arXiv:1605.04287
  [hep-ph]}}].

\bibitem{Kurkela:2018vqr}
A.~Kurkela, A.~Mazeliauskas, J.-F. Paquet, S.~Schlichting and D.~Teaney, {\it
  {Effective kinetic description of event-by-event pre-equilibrium dynamics in
  high-energy heavy-ion collisions}},
  \href{http://dx.doi.org/10.1103/PhysRevC.99.034910}{{\em Phys. Rev. C} {\bf
  99} (2019) 034910} [\href{http://arXiv.org/abs/1805.00961}{{\tt
  arXiv:1805.00961 [hep-ph]}}].

\bibitem{Lappi:2006hq}
T.~Lappi, {\it {Energy density of the glasma}},
  \href{http://dx.doi.org/10.1016/j.physletb.2006.10.017}{{\em Phys. Lett. B}
  {\bf 643} (2006) 11} [\href{http://arXiv.org/abs/hep-ph/0606207}{{\tt
  arXiv:hep-ph/0606207}}].

\bibitem{Lappi:2017skr}
T.~Lappi and S.~Schlichting, {\it {Linearly polarized gluons and axial charge
  fluctuations in the Glasma}},
  \href{http://dx.doi.org/10.1103/PhysRevD.97.034034}{{\em Phys. Rev. D} {\bf
  97} (2018) 034034} [\href{http://arXiv.org/abs/1708.08625}{{\tt
  arXiv:1708.08625 [hep-ph]}}].

\bibitem{Albacete:2018bbv}
J.~L. Albacete, P.~Guerrero-Rodr\'\i{}guez and C.~Marquet, {\it {Initial
  correlations of the Glasma energy-momentum tensor}},
  \href{http://dx.doi.org/10.1007/JHEP01(2019)073}{{\em JHEP} {\bf 01} (2019)
  073} [\href{http://arXiv.org/abs/1808.00795}{{\tt arXiv:1808.00795
  [hep-ph]}}].

\bibitem{Bhalerao:2019uzw}
R.~S. Bhalerao, G.~Giacalone, P.~Guerrero-Rodr\'\i{}guez, M.~Luzum, C.~Marquet
  and J.-Y. Ollitrault, {\it {Relating eccentricity fluctuations to density
  fluctuations in heavy-ion collisions}},
  \href{http://dx.doi.org/10.5506/APhysPolB.50.1165}{{\em Acta Phys. Polon. B}
  {\bf 50} (2019) 1165} [\href{http://arXiv.org/abs/1903.06366}{{\tt
  arXiv:1903.06366 [nucl-th]}}].

\bibitem{Giacalone:2019kgg}
G.~Giacalone, P.~Guerrero-Rodr\'\i{}guez, M.~Luzum, C.~Marquet and J.-Y.
  Ollitrault, {\it {Fluctuations in heavy-ion collisions generated by QCD
  interactions in the color glass condensate effective theory}},
  \href{http://dx.doi.org/10.1103/PhysRevC.100.024905}{{\em Phys. Rev. C} {\bf
  100} (2019) 024905} [\href{http://arXiv.org/abs/1902.07168}{{\tt
  arXiv:1902.07168 [nucl-th]}}].

\bibitem{Gelis:2019vzt}
F.~Gelis, G.~Giacalone, P.~Guerrero-Rodr\'\i{}guez, C.~Marquet and J.-Y.
  Ollitrault, {\it {Primordial fluctuations in heavy-ion collisions}},
  \href{http://arXiv.org/abs/1907.10948}{{\tt arXiv:1907.10948 [nucl-th]}}.

\bibitem{Giacalone:2019vwh}
G.~Giacalone, F.~Gelis, P.~Guerrero-Rodr\'\i{}guez, M.~Luzum, C.~Marquet and
  J.-Y. Ollitrault, {\it {Evolution of fluctuations in the initial state of
  heavy-ion collisions from RHIC to LHC}},
  \href{http://dx.doi.org/10.1007/978-3-030-53448-6_71}{{\em Springer Proc.
  Phys.} {\bf 250} (2020) 453} [\href{http://arXiv.org/abs/1911.04720}{{\tt
  arXiv:1911.04720 [nucl-th]}}].

\bibitem{Schenke:2012hg}
B.~Schenke, P.~Tribedy and R.~Venugopalan, {\it {Event-by-event gluon
  multiplicity, energy density, and eccentricities in ultrarelativistic
  heavy-ion collisions}},
  \href{http://dx.doi.org/10.1103/PhysRevC.86.034908}{{\em Phys. Rev. C} {\bf
  86} (2012) 034908} [\href{http://arXiv.org/abs/1206.6805}{{\tt
  arXiv:1206.6805 [hep-ph]}}].

\bibitem{Schenke:2012wb}
B.~Schenke, P.~Tribedy and R.~Venugopalan, {\it {Fluctuating Glasma initial
  conditions and flow in heavy ion collisions}},
  \href{http://dx.doi.org/10.1103/PhysRevLett.108.252301}{{\em Phys. Rev.
  Lett.} {\bf 108} (2012) 252301} [\href{http://arXiv.org/abs/1202.6646}{{\tt
  arXiv:1202.6646 [nucl-th]}}].

\bibitem{Mantysaari:2017cni}
H.~M\"antysaari, B.~Schenke, C.~Shen and P.~Tribedy, {\it {Imprints of
  fluctuating proton shapes on flow in proton-lead collisions at the LHC}},
  \href{http://dx.doi.org/10.1016/j.physletb.2017.07.038}{{\em Phys. Lett. B}
  {\bf 772} (2017) 681} [\href{http://arXiv.org/abs/1705.03177}{{\tt
  arXiv:1705.03177 [nucl-th]}}].

\bibitem{Jia_2021}
M.~Jia, J.~Liu, H.~Zhang and M.~Ruggieri, {\it Fluctuations of topological
  charge and chiral density in the early stage of high energy nuclear
  collisions},  \href{http://dx.doi.org/10.1103/physrevd.103.014026}{{\em
  Physical Review D} {\bf 103} (2021) }[\href{http://arxiv.org/abs/2006.01090}{{\tt
  arXiv:2006.01090 [hep-ph]}}].

\bibitem{Krasnitz:1998ns}
A.~Krasnitz and R.~Venugopalan, {\it {Nonperturbative computation of gluon
  minijet production in nuclear collisions at very high-energies}},
  \href{http://dx.doi.org/10.1016/S0550-3213(99)00366-1}{{\em Nucl. Phys. B}
  {\bf 557} (1999) 237} [\href{http://arXiv.org/abs/hep-ph/9809433}{{\tt
  arXiv:hep-ph/9809433}}].

\bibitem{Krasnitz:2001qu}
A.~Krasnitz, Y.~Nara and R.~Venugopalan, {\it {Coherent gluon production in
  very high-energy heavy ion collisions}},
  \href{http://dx.doi.org/10.1103/PhysRevLett.87.192302}{{\em Phys. Rev. Lett.}
  {\bf 87} (2001) 192302} [\href{http://arXiv.org/abs/hep-ph/0108092}{{\tt
  arXiv:hep-ph/0108092}}].

\bibitem{Kovner:1995ts}
A.~Kovner, L.~D. McLerran and H.~Weigert, {\it {Gluon production at high
  transverse momentum in the McLerran-Venugopalan model of nuclear structure
  functions}},  \href{http://dx.doi.org/10.1103/PhysRevD.52.3809}{{\em Phys.
  Rev. D} {\bf 52} (1995) 3809}
  [\href{http://arXiv.org/abs/hep-ph/9505320}{{\tt arXiv:hep-ph/9505320}}].

\bibitem{Kovchegov:1997ke}
Y.~V. Kovchegov and D.~H. Rischke, {\it {Classical gluon radiation in
  ultrarelativistic nucleus-nucleus collisions}},
  \href{http://dx.doi.org/10.1103/PhysRevC.56.1084}{{\em Phys. Rev. C} {\bf 56}
  (1997) 1084} [\href{http://arXiv.org/abs/hep-ph/9704201}{{\tt
  arXiv:hep-ph/9704201}}].

\bibitem{Dumitru:2001ux}
A.~Dumitru and L.~D. McLerran, {\it {How protons shatter colored glass}},
  \href{http://dx.doi.org/10.1016/S0375-9474(01)01301-X}{{\em Nucl. Phys. A}
  {\bf 700} (2002) 492} [\href{http://arXiv.org/abs/hep-ph/0105268}{{\tt
  arXiv:hep-ph/0105268}}].

\bibitem{McLerran:2016snu}
L.~McLerran and V.~Skokov, {\it {Odd Azimuthal Anisotropy of the Glasma for pA
  Scattering}},  \href{http://dx.doi.org/10.1016/j.nuclphysa.2016.12.011}{{\em
  Nucl. Phys. A} {\bf 959} (2017) 83}
  [\href{http://arXiv.org/abs/1611.09870}{{\tt arXiv:1611.09870 [hep-ph]}}].

\bibitem{Fries:2006pv}
R.~J. Fries, J.~I. Kapusta and Y.~Li, {\it {Near-fields and initial energy
  density in the color glass condensate model}},
  \href{http://arXiv.org/abs/nucl-th/0604054}{{\tt arXiv:nucl-th/0604054}}.

\bibitem{Fujii:2008km}
H.~Fujii, K.~Fukushima and Y.~Hidaka, {\it {Initial energy density and gluon
  distribution from the Glasma in heavy-ion collisions}},
  \href{http://dx.doi.org/10.1103/PhysRevC.79.024909}{{\em Phys. Rev. C} {\bf
  79} (2009) 024909} [\href{http://arXiv.org/abs/0811.0437}{{\tt
  arXiv:0811.0437 [hep-ph]}}].

\bibitem{Chen:2015wia}
G.~Chen, R.~J. Fries, J.~I. Kapusta and Y.~Li, {\it {Early Time Dynamics of
  Gluon Fields in High Energy Nuclear Collisions}},
  \href{http://dx.doi.org/10.1103/PhysRevC.92.064912}{{\em Phys. Rev. C} {\bf
  92} (2015) 064912} [\href{http://arXiv.org/abs/1507.03524}{{\tt
  arXiv:1507.03524 [nucl-th]}}].

\bibitem{Carrington:2020ssh}
M.~E. Carrington, A.~Czajka and S.~Mrowczynski, {\it {The energy-momentum
  tensor at the earliest stage of relativistic heavy ion collisions --
  formalism}},  \href{http://arXiv.org/abs/2012.03042}{{\tt arXiv:2012.03042
  [hep-ph]}}.

\bibitem{Blaizot:2010kh}
J.~P. Blaizot, T.~Lappi and Y.~Mehtar-Tani, {\it {On the gluon spectrum in the
  glasma}},  \href{http://dx.doi.org/10.1016/j.nuclphysa.2010.06.009}{{\em
  Nucl. Phys. A} {\bf 846} (2010) 63}
  [\href{http://arXiv.org/abs/1005.0955}{{\tt arXiv:1005.0955 [hep-ph]}}].

\bibitem{Boguslavski:2021buh}
K.~Boguslavski, A.~Kurkela, T.~Lappi and J.~Peuron, {\it {Broad excitations in
  a 2+1D overoccupied gluon plasma}}, \href{http://dx.doi.org/10.1007/JHEP05(2021)225}{\em JHEP {\bf 05} (2021) 225}
  [\href{http://arXiv.org/abs/2101.02715}{{\tt arXiv:2101.02715 [hep-ph]}}].

\bibitem{Boguslavski:2019fsb}
K.~Boguslavski, A.~Kurkela, T.~Lappi and J.~Peuron, {\it {Highly occupied gauge
  theories in 2+1 dimensions: A self-similar attractor}},
  \href{http://dx.doi.org/10.1103/PhysRevD.100.094022}{{\em Phys. Rev. D} {\bf
  100} (2019) 094022} [\href{http://arXiv.org/abs/1907.05892}{{\tt
  arXiv:1907.05892 [hep-ph]}}].

\bibitem{JalilianMarian:1997gr}
J.~Jalilian-Marian, A.~Kovner, A.~Leonidov and H.~Weigert, {\it {The Wilson
  renormalization group for low x physics: Towards the high density regime}},
  \href{http://dx.doi.org/10.1103/PhysRevD.59.014014}{{\em Phys. Rev. D} {\bf
  59} (1998) 014014} [\href{http://arXiv.org/abs/hep-ph/9706377}{{\tt
  arXiv:hep-ph/9706377}}].

\bibitem{JalilianMarian:1997dw}
J.~Jalilian-Marian, A.~Kovner and H.~Weigert, {\it {The Wilson renormalization
  group for low x physics: Gluon evolution at finite parton density}},
  \href{http://dx.doi.org/10.1103/PhysRevD.59.014015}{{\em Phys. Rev. D} {\bf
  59} (1998) 014015} [\href{http://arXiv.org/abs/hep-ph/9709432}{{\tt
  arXiv:hep-ph/9709432}}].

\bibitem{Weigert:2000gi}
H.~Weigert, {\it {Unitarity at small Bjorken x}},
  \href{http://dx.doi.org/10.1016/S0375-9474(01)01668-2}{{\em Nucl. Phys. A}
  {\bf 703} (2002) 823} [\href{http://arXiv.org/abs/hep-ph/0004044}{{\tt
  arXiv:hep-ph/0004044}}].

\bibitem{Iancu:2000hn}
E.~Iancu, A.~Leonidov and L.~D. McLerran, {\it {Nonlinear gluon evolution in
  the color glass condensate. 1.}},
  \href{http://dx.doi.org/10.1016/S0375-9474(01)00642-X}{{\em Nucl. Phys. A}
  {\bf 692} (2001) 583} [\href{http://arXiv.org/abs/hep-ph/0011241}{{\tt
  arXiv:hep-ph/0011241}}].

\bibitem{Ferreiro:2001qy}
E.~Ferreiro, E.~Iancu, A.~Leonidov and L.~McLerran, {\it {Nonlinear gluon
  evolution in the color glass condensate. 2.}},
  \href{http://dx.doi.org/10.1016/S0375-9474(01)01329-X}{{\em Nucl. Phys. A}
  {\bf 703} (2002) 489} [\href{http://arXiv.org/abs/hep-ph/0109115}{{\tt
  arXiv:hep-ph/0109115}}].

\bibitem{Balitsky:1995ub}
I.~Balitsky, {\it {Operator expansion for high-energy scattering}},
  \href{http://dx.doi.org/10.1016/0550-3213(95)00638-9}{{\em Nucl. Phys. B}
  {\bf 463} (1996) 99} [\href{http://arXiv.org/abs/hep-ph/9509348}{{\tt
  arXiv:hep-ph/9509348}}].

\bibitem{Kovchegov:1999yj}
Y.~V. Kovchegov, {\it {Small x F(2) structure function of a nucleus including
  multiple pomeron exchanges}},
  \href{http://dx.doi.org/10.1103/PhysRevD.60.034008}{{\em Phys. Rev. D} {\bf
  60} (1999) 034008} [\href{http://arXiv.org/abs/hep-ph/9901281}{{\tt
  arXiv:hep-ph/9901281}}].

\bibitem{Metz:2011wb}
A.~Metz and J.~Zhou, {\it {Distribution of linearly polarized gluons inside a
  large nucleus}},  \href{http://dx.doi.org/10.1103/PhysRevD.84.051503}{{\em
  Phys. Rev. D} {\bf 84} (2011) 051503}
  [\href{http://arXiv.org/abs/1105.1991}{{\tt arXiv:1105.1991 [hep-ph]}}].

\bibitem{GolecBiernat:1998js}
K.~J. Golec-Biernat and M.~Wusthoff, {\it {Saturation effects in deep inelastic
  scattering at low Q**2 and its implications on diffraction}},
  \href{http://dx.doi.org/10.1103/PhysRevD.59.014017}{{\em Phys. Rev. D} {\bf
  59} (1998) 014017} [\href{http://arXiv.org/abs/hep-ph/9807513}{{\tt
  arXiv:hep-ph/9807513}}].

\bibitem{Lappi:2007ku}
T.~Lappi, {\it {Wilson line correlator in the MV model: Relating the glasma to
  deep inelastic scattering}},
  \href{http://dx.doi.org/10.1140/epjc/s10052-008-0588-4}{{\em Eur. Phys. J. C}
  {\bf 55} (2008) 285} [\href{http://arXiv.org/abs/0711.3039}{{\tt
  arXiv:0711.3039 [hep-ph]}}].

\bibitem{JalilianMarian:1996xn}
J.~Jalilian-Marian, A.~Kovner, L.~D. McLerran and H.~Weigert, {\it {The
  Intrinsic glue distribution at very small x}},
  \href{http://dx.doi.org/10.1103/PhysRevD.55.5414}{{\em Phys. Rev. D} {\bf 55}
  (1997) 5414} [\href{http://arXiv.org/abs/hep-ph/9606337}{{\tt
  arXiv:hep-ph/9606337}}].

\bibitem{Dusling:2009ni}
K.~Dusling, F.~Gelis, T.~Lappi and R.~Venugopalan, {\it {Long range
  two-particle rapidity correlations in A+A collisions from high energy QCD
  evolution}},  \href{http://dx.doi.org/10.1016/j.nuclphysa.2009.12.044}{{\em
  Nucl. Phys. A} {\bf 836} (2010) 159}
  [\href{http://arXiv.org/abs/0911.2720}{{\tt arXiv:0911.2720 [hep-ph]}}].

\bibitem{Dumitru:2010iy}
A.~Dumitru, K.~Dusling, F.~Gelis, J.~Jalilian-Marian, T.~Lappi and
  R.~Venugopalan, {\it {The Ridge in proton-proton collisions at the LHC}},
  \href{http://dx.doi.org/10.1016/j.physletb.2011.01.024}{{\em Phys. Lett. B}
  {\bf 697} (2011) 21} [\href{http://arXiv.org/abs/1009.5295}{{\tt
  arXiv:1009.5295 [hep-ph]}}].

\bibitem{Dusling:2012iga}
K.~Dusling and R.~Venugopalan, {\it {Azimuthal collimation of long range
  rapidity correlations by strong color fields in high multiplicity
  hadron-hadron collisions}},
  \href{http://dx.doi.org/10.1103/PhysRevLett.108.262001}{{\em Phys. Rev.
  Lett.} {\bf 108} (2012) 262001} [\href{http://arXiv.org/abs/1201.2658}{{\tt
  arXiv:1201.2658 [hep-ph]}}].

\bibitem{Dusling:2012cg}
K.~Dusling and R.~Venugopalan, {\it {Evidence for BFKL and saturation dynamics
  from dihadron spectra at the LHC}},
  \href{http://dx.doi.org/10.1103/PhysRevD.87.051502}{{\em Phys. Rev. D} {\bf
  87} (2013) 051502} [\href{http://arXiv.org/abs/1210.3890}{{\tt
  arXiv:1210.3890 [hep-ph]}}].

\bibitem{Dusling:2012wy}
K.~Dusling and R.~Venugopalan, {\it {Explanation of systematics of CMS p+Pb
  high multiplicity di-hadron data at $\sqrt{s}_{\rm NN} = 5.02$ TeV}},
  \href{http://dx.doi.org/10.1103/PhysRevD.87.054014}{{\em Phys. Rev. D} {\bf
  87} (2013) 054014} [\href{http://arXiv.org/abs/1211.3701}{{\tt
  arXiv:1211.3701 [hep-ph]}}].

\bibitem{Mertig:1990an}
R.~Mertig, M.~Bohm and A.~Denner, {\it {FEYN CALC: Computer algebraic
  calculation of Feynman amplitudes}},
  \href{http://dx.doi.org/10.1016/0010-4655(91)90130-D}{{\em Comput. Phys.
  Commun.} {\bf 64} (1991) 345}.

\bibitem{Shtabovenko:2016sxi}
V.~Shtabovenko, R.~Mertig and F.~Orellana, {\it {New Developments in FeynCalc
  9.0}},  \href{http://dx.doi.org/10.1016/j.cpc.2016.06.008}{{\em Comput. Phys.
  Commun.} {\bf 207} (2016) 432} [\href{http://arXiv.org/abs/1601.01167}{{\tt
  arXiv:1601.01167 [hep-ph]}}].

\bibitem{Boguslavski:2018beu}
K.~Boguslavski, A.~Kurkela, T.~Lappi and J.~Peuron, {\it {Spectral function for
  overoccupied gluodynamics from real-time lattice simulations}},
  \href{http://dx.doi.org/10.1103/PhysRevD.98.014006}{{\em Phys. Rev. D} {\bf
  98} (2018) 014006} [\href{http://arXiv.org/abs/1804.01966}{{\tt
  arXiv:1804.01966 [hep-ph]}}].

\bibitem{Kovchegov:2001sc}
Y.~V. Kovchegov and K.~Tuchin, {\it {Inclusive gluon production in DIS at high
  parton density}},  \href{http://dx.doi.org/10.1103/PhysRevD.65.074026}{{\em
  Phys. Rev. D} {\bf 65} (2002) 074026}
  [\href{http://arXiv.org/abs/hep-ph/0111362}{{\tt arXiv:hep-ph/0111362}}].

\bibitem{Blaizot:2004wu}
J.~P. Blaizot, F.~Gelis and R.~Venugopalan, {\it {High-energy pA collisions in
  the color glass condensate approach. 1. Gluon production and the Cronin
  effect}},  \href{http://dx.doi.org/10.1016/j.nuclphysa.2004.07.005}{{\em
  Nucl. Phys. A} {\bf 743} (2004) 13}
  [\href{http://arXiv.org/abs/hep-ph/0402256}{{\tt arXiv:hep-ph/0402256}}].

\bibitem{Li:2021zmf}
M.~Li and V.~V. Skokov, {\it {First Saturation Correction in High Energy
  Proton-Nucleus Collisions: I. Time evolution of classical Yang-Mills fields
  beyond leading order}}, \href{http://dx.doi.org/10.1007/JHEP06(2021)140}{{\em
  JHEP} {\bf 06} (2021) 140}
  [\href{http://arXiv.org/abs/2102.01594}{{\tt arXiv:2102.01594 [hep-ph]}}].

\bibitem{Kovchegov:2005ss}
Y.~V. Kovchegov, {\it {Can thermalization in heavy ion collisions be described
  by QCD diagrams?}},
  \href{http://dx.doi.org/10.1016/j.nuclphysa.2005.08.009}{{\em Nucl. Phys. A}
  {\bf 762} (2005) 298} [\href{http://arXiv.org/abs/hep-ph/0503038}{{\tt
  arXiv:hep-ph/0503038}}].

\bibitem{Kovchegov:2005kn}
Y.~V. Kovchegov, {\it {Thoughts on non-perturbative thermalization and jet
  quenching in heavy ion collisions}},
  \href{http://dx.doi.org/10.1016/j.nuclphysa.2005.09.010}{{\em Nucl. Phys. A}
  {\bf 764} (2006) 476} [\href{http://arXiv.org/abs/hep-ph/0507134}{{\tt
  arXiv:hep-ph/0507134}}].

\bibitem{Fries:2007iy}
R.~J. Fries, {\it {Early Time Evolution of High Energy Heavy Ion Collisions}},
  \href{http://dx.doi.org/10.1088/0954-3899/34/8/S111}{{\em J. Phys. G} {\bf
  34} (2007) S851} [\href{http://arXiv.org/abs/nucl-th/0702026}{{\tt
  arXiv:nucl-th/0702026}}].

\bibitem{Skokov:2009qp}
V.~Skokov, A.~Y. Illarionov and V.~Toneev, {\it {Estimate of the magnetic field
  strength in heavy-ion collisions}},
  \href{http://dx.doi.org/10.1142/S0217751X09047570}{{\em Int. J. Mod. Phys. A}
  {\bf 24} (2009) 5925} [\href{http://arXiv.org/abs/0907.1396}{{\tt
  arXiv:0907.1396 [nucl-th]}}].

\bibitem{Kharzeev:2001ev}
D.~Kharzeev, A.~Krasnitz and R.~Venugopalan, {\it {Anomalous chirality
  fluctuations in the initial stage of heavy ion collisions and parity odd
  bubbles}},  \href{http://dx.doi.org/10.1016/S0370-2693(02)02630-8}{{\em Phys.
  Lett. B} {\bf 545} (2002) 298}
  [\href{http://arXiv.org/abs/hep-ph/0109253}{{\tt arXiv:hep-ph/0109253}}].

\bibitem{Kharzeev:2007jp}
D.~E. Kharzeev, L.~D. McLerran and H.~J. Warringa, {\it {The Effects of
  topological charge change in heavy ion collisions: 'Event by event P and CP
  violation'}},  \href{http://dx.doi.org/10.1016/j.nuclphysa.2008.02.298}{{\em
  Nucl. Phys. A} {\bf 803} (2008) 227}
  [\href{http://arXiv.org/abs/0711.0950}{{\tt arXiv:0711.0950 [hep-ph]}}].

\bibitem{Voloshin:2004vk}
S.~A. Voloshin, {\it {Parity violation in hot QCD: How to detect it}},
  \href{http://dx.doi.org/10.1103/PhysRevC.70.057901}{{\em Phys. Rev. C} {\bf
  70} (2004) 057901} [\href{http://arXiv.org/abs/hep-ph/0406311}{{\tt
  arXiv:hep-ph/0406311}}].

\bibitem{Bzdak:2012ia}
A.~Bzdak, V.~Koch and J.~Liao, {\it {Charge-Dependent Correlations in
  Relativistic Heavy Ion Collisions and the Chiral Magnetic Effect}},
  \href{http://dx.doi.org/10.1007/978-3-642-37305-3_19}{{\em Lect. Notes Phys.}
  {\bf 871} (2013) 503} [\href{http://arXiv.org/abs/1207.7327}{{\tt
  arXiv:1207.7327 [nucl-th]}}].

\bibitem{Wen:2016zic}
F.~Wen, J.~Bryon, L.~Wen and G.~Wang, {\it {Event-shape-engineering study of
  charge separation in heavy-ion collisions}},
  \href{http://dx.doi.org/10.1088/1674-1137/42/1/014001}{{\em Chin. Phys. C}
  {\bf 42} (2018) 014001} [\href{http://arXiv.org/abs/1608.03205}{{\tt
  arXiv:1608.03205 [nucl-th]}}].

\bibitem{Xu:2017qfs}
H.-j. Xu, J.~Zhao, X.~Wang, H.~Li, Z.-W. Lin, C.~Shen and F.~Wang, {\it
  {Varying the chiral magnetic effect relative to flow in a single
  nucleus-nucleus collision}},
  \href{http://dx.doi.org/10.1088/1674-1137/42/8/084103}{{\em Chin. Phys. C}
  {\bf 42} (2018) 084103} [\href{http://arXiv.org/abs/1710.07265}{{\tt
  arXiv:1710.07265 [nucl-th]}}].

\bibitem{Huang:2018duz}
H.~Huang {\em et.~al.}, {\it {Operation of RHIC injectors with isobaric Ruthenium
  and Zirconium ions}} in \href{http://doi.org/10.18429/JACoW-IPAC2018-TUPAF006}{{\em {9th International Particle Accelerator
  Conference}}}, (JACoW Publishing,
Geneva, Switzerland, 2018), p.~TUPAF006, 2018.

\bibitem{Schlichting:2010qia}
S.~Schlichting and S.~Pratt, {\it {Charge conservation at energies available at
  the BNL Relativistic Heavy Ion Collider and contributions to local parity
  violation observables}},
  \href{http://dx.doi.org/10.1103/PhysRevC.83.014913}{{\em Phys. Rev. C} {\bf
  83} (2011) 014913} [\href{http://arXiv.org/abs/1009.4283}{{\tt
  arXiv:1009.4283 [nucl-th]}}].

\bibitem{Guerrero-Rodriguez:2019ids}
P.~Guerrero-Rodr\'\i{}guez, {\it {Topological charge fluctuations in the
  Glasma}},  \href{http://dx.doi.org/10.1007/JHEP08(2019)026}{{\em JHEP} {\bf
  08} (2019) 026} [\href{http://arXiv.org/abs/1903.11602}{{\tt arXiv:1903.11602
  [hep-ph]}}].

\bibitem{Hammelmann:2019vwd}
J.~Hammelmann, A.~Soto-Ontoso, M.~Alvioli, H.~Elfner and M.~Strikman, {\it
  {Influence of the neutron-skin effect on nuclear isobar collisions at
  energies available at the BNL Relativistic Heavy Ion Collider}},
  \href{http://dx.doi.org/10.1103/PhysRevC.101.061901}{{\em Phys. Rev. C} {\bf
  101} (2020) 061901} [\href{http://arXiv.org/abs/1908.10231}{{\tt
  arXiv:1908.10231 [nucl-th]}}].

\end{thebibliography}
\providecommand{\href}[2]{#2}\begingroup\raggedright\endgroup

\end{document}